\documentclass[iop]{emulateapj}

\usepackage{amsmath}
\usepackage{amssymb}
\usepackage{color}
\usepackage{epic}
\usepackage{epstopdf}
\usepackage{graphicx}
\usepackage{hyperref}
\usepackage{multirow}
\usepackage{natbib}
\usepackage{overpic}
\usepackage{slashbox}

\shortauthors{Kapi\'nska et al.}
\shorttitle{RGZ HyMoRS Candidates}

\begin{document}

\title{Radio Galaxy Zoo: A Search for Hybrid morphology radio galaxies}

\author{A. D. Kapi\'{n}ska$^{1,2}$, I. Terentev$^3$, O. I. Wong$^{1,2}$, S. S. Shabala$^4$, H. Andernach$^5$, L.~Rudnick$^6$, L. Storer$^1$,  J.~K.~Banfield$^{2,7}$,  K.~W.~Willett$^6$, F. de Gasperin$^8$, C.~J.~Lintott$^{9}$, \'A.~R. L\'opez-S\'anchez$^{10,11}$, E. Middelberg$^{12}$, R. P. Norris$^{2,13,14}$,  K. Schawinski$^{15}$, N. Seymour$^{16}$, B. Simmons$^{9,17}$}
\email{\tt E-mail: anna.kapinska@uwa.edu.au}

\affil{
$^1$ International Centre for Radio Astronomy Research (ICRAR), University of Western Australia, 35 Stirling Hwy, WA 6009, Australia\\
$^2$ ARC Centre of Excellence for All-Sky Astrophysics (CAASTRO)\\
$^3$ Zooniverse Citizen Scientist, c/o Oxford Astrophysics, Denys Wilkinson Building, Keble Road, Oxford OX1 3RH, U.K.\\
$^4$ School of Physical Sciences, University of Tasmania, Private Bag 37, Hobart, Tasmania 7001, Australia  \\
$^5$ Depto. de Astronom\'{i}a, DCNE, Universidad de Guanajuato, Apdo. Postal 144, Guanajuato, CP 36000, Gto., Mexico\\
$^6$ School of Physics and Astronomy, University of Minnesota, Minneapolis, MN 55455, U.S.A.\\
$^7$ Research School of Astronomy and Astrophysics, Australian National University, Canberra, ACT 2611, Australia\\
$^8$ Leiden Observatory, Leiden University, P.O.Box 9513, NL-2300 RA, Leiden, The Netherlands\\
$^{9}$ Astrophysics, Department of Physics, University of Oxford, Keble Road, Oxford OX1 3RH, U.K.\\
$^{10}$ Australian Astronomical Observatory, PO Box 915, North Ryde, NSW 1670, Australia\\
$^{11}$ Department of Physics and Astronomy, Macquarie University, NSW 2109, Australia\\
$^{12}$ Astronomisches Institut, Ruhr-Universit\"at, Universit\"atsstr. 150, 44801 Bochum, Germany\\
$^{13}$ CSIRO Australia Telescope National Facility, PO Box 76, Epping, NSW 1710, Australia\\
$^{14}$ Western Sydney University, Locked Bag 1797, Penrith South, NSW 1797, Australia\\
$^{15}$ Institute for Astronomy, Department of Physics, ETH Zurich, Wolfgang-Pauli-Strasse 27, CH-8093 Zurich, Switzerland\\
$^{16}$ International Centre for Radio Astronomy Research (ICRAR), Curtin University, Perth, Australia  \\
$^{17}$ {Center for Astrophysics and Space Sciences, Department of Physics, University of California, San Diego, CA 92093, U.S.A.}
}

\begin{abstract}

Hybrid morphology radio sources are a rare type of radio galaxy that display different Fanaroff-Riley classes on opposite sides of their nuclei. To enhance the statistical analysis of hybrid morphology radio sources, we embarked on a large-scale search of these sources within the international citizen science project, Radio Galaxy Zoo (RGZ). Here, we present 25 new candidate hybrid morphology radio galaxies. Our selected candidates are moderate power radio galaxies ($L_{\rm median} = {4.7}\times 10^{24}$~W~Hz$^{-1}$~sr$^{-1}$) at redshifts ${0.14} < z < 1.0$. Hosts of {nine} candidates have spectroscopic observations, of which {six} are classified as quasars, one as high- and two as low-excitation galaxies. Two candidate HyMoRS are giant ($>1$~Mpc) radio galaxies, one resides {at} a center of a galaxy cluster, and one is hosted by a rare green bean galaxy. Although the origin of the hybrid morphology radio galaxies is still unclear, this type of radio source starts depicting itself as a rather diverse class. We discuss hybrid radio morphology formation in terms of  the radio source environment (nurture) and intrinsically occurring phenomena (nature; activity cessation and amplification), showing that these peculiar radio galaxies can be formed by both mechanisms. While high angular resolution follow-up observations are still necessary to confirm our candidates, we demonstrate the efficacy of the Radio Galaxy Zoo in the pre-selection of these sources from all-sky radio surveys, and report  the reliability of citizen scientists in identifying and classifying complex radio sources. 

\end{abstract}



\section{Introduction}
\label{sec:intro}

Hybrid morphology radio sources (hereafter HyMoRS) have been invoked in the debate on the origin of morphological dichotomy { in} radio galaxies since \cite{2000A&A...363..507G} pointed out their existence. \cite{2000A&A...363..507G} claimed that HyMoRS constitute a separate class of object alongside the large-scale radio galaxies of FR~I and FR~II types \cite[][]{1974MNRAS.167P..31F}; the Fanaroff-Riley classes (FR) introduced over 40 years ago have proved to be a simple yet powerful tool in studies of radio galaxies.

In the original definition FR~I and FR~II radio sources were distinguished by the ratio, $\mathcal{R}$, of the distance between the brightness peaks of each side of the nucleus {and} the overall source size. If $\mathcal{R}<0.5$ the radio source was classified as an FR~I, otherwise as an FR~II. Since the formulation of the FR classification our understanding of radio galaxies has been evolving, both observationally and theoretically, and we now know that FR~II radio galaxies exhibit tightly collimated and remarkably stable, often one-sided jets, which terminate forming well recognised  features, the so-called hotspots. It is interpreted that the relativistic particles \cite[either electron/proton or electron/positron pairs, see e.g.][]{2006ApJ...648..200D} are re-accelerated in strong shocks at these jet termination points, and are further transported, through the backflow, into a cocoon encompassing the radio source \cite[][]{1974MNRAS.169..395B,1974MNRAS.166..513S}. FR~Is, on the other hand, seem to display heavily turbulent jets; they are less well collimated and have been shown to decelerate while strongly interacting with the ambient medium soon after their ejection \cite[][]{1988Ap&SS.150...59K,2014MNRAS.437.3405L}.

The physical origin of the FR dichotomy in radio galaxy population has been a widely debated issue for over 20 years \cite[e.g.][among others]{1992ApJ...389..208B,1995ApJ...451...88B,1993ApJ...405L..13D,1996MNRAS.283L.111R,1997Natur.388..350M,2002IAUS..199...34R,2007MNRAS.381.1548K,2009ApJ...697L.173K,2012AJ....144...85S,2013MNRAS.430.3086G}. Theoretical works \cite[][]{1994AuJPh..47..669B,1995ApJS..101...29B,2007MNRAS.381.1548K,2009ApJ...697L.173K} explain the dichotomy in terms of power of the relativistic outflows, { and} deceleration and interaction of the jets with their environment. Such an interpretation has been supported by a wealth of observational studies \cite[e.g.][]{1991ApJ...367....1H,1997ApJ...476..489Z,2011MNRAS.417.2789L,2013MNRAS.430.3086G,2013MNRAS.434.2877T,2014MNRAS.437.3405L}. Also, analytical solutions that link the transition of FR~II sources into FR~Is have been developed \cite[][]{2011MNRAS.418.1138W,2015Turner}. On the other hand, some authors favor fundamental differences in the central engines of the two different FR classes as the interpretation of the dichotomy. For instance, through spectroscopic analysis of emission line nebulae associated with radio galaxies, \cite{1992ApJ...389..208B,1995ApJ...451...88B} suggested that the angular momentum of the accretion disk may be important in forming radio morphologies of extragalactic radio sources. Baum et al. suggested that the AGN in FR~I radio galaxies are fed at low accretion rates and are possibly of low black hole spins, while those in FR~II radio galaxies are expected to accrete at higher rates and be possibly of higher spins. Fundamental differences in the central engines of the two different FR classes are also favored by e.g. \cite{1997Natur.388..350M} and \cite{2002IAUS..199...34R}. Differences in the particle composition of the jets have also been considered \cite[e.g.][]{1993MNRAS.264..228C,1996MNRAS.283..873R,2002MNRAS.336.1161L}. 

The current consensus is that the FR morphology is most likely due to a combination of jet power and the radio source environment \cite[e.g.][]{2009AN....330..184B,2012AJ....144...85S}. HyMoRS, which -- in simple terms -- display FR~I radio structures on one side of the nucleus and FR~II on the other, seem to be a class of object that may help us to disentangle the two effects. However, HyMoRS are a rare type of radio galaxy, with an estimated occurrence possibly as low as $<1$\% of all extended radio galaxies \cite[][]{2006A&A...447...63G}. Here, we demonstrate how the Radio Galaxy Zoo project \cite[RGZ;][]{2015MNRAS.453.2326B} is particularly useful in the search for HyMoRS. 

This paper is structured as follows. We present the project and our data in \S\ref{sec:RGZ}. Our results, including notes { on the RGZ} candidate HyMoRS, their optical and radio properties, and performance of the RGZ project in classifying HyMoRS, are reported in \S\ref{sec:results}. We discuss the possible origin of HyMoRS, their host galaxies, environments and radio properties in \S\ref{sec:disc}. Summary and conclusions are given in \S\ref{sec:summary}. We assume a flat Universe with the Hubble constant of $H_0 = 68$ km~s$^{-1}$~Mpc$^{-1}$, and $\Omega_{\rm \Lambda} = 0.685$ and $\Omega_{\rm M} = 0.315$ \cite[][]{2014A&A...571A..16P} throughout the paper.


\section{Observational Data}
\label{sec:RGZ}

Our source of data for this work is the Radio Galaxy Zoo\footnote{\tt http://radio.galaxyzoo.org/} citizen science project  \cite[][]{2015MNRAS.453.2326B}. RGZ builds on the hugely successful Galaxy Zoo\footnote{\tt http://www.galaxyzoo.org/} \cite[][]{2008MNRAS.389.1179L}. RGZ uses major legacy radio and mid-IR large area surveys: the Faint Images of the Radio Sky at Twenty centimeters \cite[FIRST;][]{1995ApJ...450..559B}, the Australia Telescope Large Area Survey \cite[ATLAS; ][]{2006AJ....132.2409N}, the {\it Wide-field Infrared Survey Explorer} \cite[{\it WISE};][]{2010AJ....140.1868W}, and the {\it Spitzer} Wide-Area Infrared Extragalactic Survey \cite[SWIRE; ][]{2003PASP..115..897L}.

\subsection{Radio Galaxy Zoo Talk}

In this paper we present a sample of candidate hybrid morphology radio galaxies serendipitously identified by the RGZ citizen scientists, and discussed within the RGZ online forum, the RadioTalk\footnote{\tt http://radiotalk.galaxyzoo.org/}. In the standard process, the RGZ citizen scientists are provided with  $3'\times3'$ cut-outs drawn from the radio and mid-IR surveys that present them with a `subject' to classify. Their task is to (i) decide whether radio components in the FIRST cut-out are separate radio sources, or if they belong to one galaxy, and (ii) locate the mid-IR host galaxy of the radio source(s), if present, in the corresponding {\it WISE} image \cite[for details see][]{2015MNRAS.453.2326B}. After the classification of each cut-out the citizen scientists can discuss the subjects they have just classified with the RGZ science team members and other volunteers through the RadioTalk forum. We follow up, with detailed visual inspection, radio sources tagged as { having} `hybrid' and `asymmetric' radio morphology and discussed by the volunteers on RadioTalk. 
In total, we inspected visually 427 sources, and the 25 best cases, all of which are reported in this paper, are considered strong HyMoRS candidates. However, the current sample cannot be used to give a quantitative estimate of the fraction of sources which fall into the HyMoRS category, because of the non-uniform way in which the sources are found. The pre-selection of the sources for RadioTalk by the volunteers is incidental, there is an unknown bias associated with which sources were tagged as hybrid or asymmetric, and a number of our candidates were found serendipitously by the science team.


\begin{table*}
\begin{minipage}{17.5cm}
\footnotesize
\centering
\caption{Optical hosts of candidate RGZ HyMoRS. Redshifts, optical positions, and $r$-band { cmodel} magnitudes, {corrected for Galactic extinction and transformed into R-band \protect\cite[][]{2005AJ....130..873J}}, are sourced from the SDSS DR13 unless stated otherwise.}
\label{tab:optical}
\begin{tabular}{lrrlllllcl}
\hline
\hline
HyMoRS name              & RA (J2000) & Dec (J2000) & \multicolumn{1}{c}{$m_{\rm r}$} &  \multicolumn{1}{c}{$m_{\rm R}$}  &  \multicolumn{1}{c}{$M_{\rm R}$}  &  \multicolumn{1}{c}{$z$} &  \multicolumn{1}{c}{$z_{\rm err}$} & \multicolumn{1}{c}{$z_{\rm type}$}  &  \\
\hline
RGZ J023832.{6}+023349   & 02 38 32.67  & +02 33 49.2   & $17.1{4}\pm0.01$ & ${17.06\pm0.02}$ & ${-22.86\pm0.02}$   & {0.209}     & { 0.001}    & {spectro$^{\dagger}$}\\
RGZ J072406.{7}+38034{8} & 07 24 06.79  & +38 03 48.6   & $17.{70}\pm0.01$ & ${17.70\pm0.02}$ & ${-22.31\pm0.02}$   & $0.2414$    & $0.00002$   & spectro \\
RGZ J082231.{0}+53111{8} & 08 22 34.06  & +53 11 18.7   & $18.{18\pm0.04}$ & ${18.18\pm0.07}$ & ${-20.81\pm0.08}^*$ & $0.138$     & $0.011$     & photo \\
RGZ J083352.2+04582{2}   & 08 33 52.25  & +04 58 22.4   & $19.{58\pm0.03}$ & ${19.82\pm0.24}$ & ${-20.31\pm0.32}^*$ & {0.227}     & {0.114}     & photo{ $^{\ddagger}$} \\
RGZ J084738.{0}+183156   & -- & -- & -- & -- & -- & --& -- & --\\
RGZ J085926.7+292738     & 08 59 26.71  & +29 27 38.1   & $18.{86}\pm0.02$ & ${19.53\pm0.05}$ & ${-20.87\pm0.07}$    & $0.272$     & $0.021$     & photo \\
RGZ J091408.0+52294{8}   & 09 14 08.01  & +52 29 48.6   & ${20.97\pm0.11}$ & ${20.97\pm0.27}$ & ${-20.86\pm0.28}$    & $0.607$     & $0.039$     & photo \\
RGZ J094214.2+06275{2}   & 09 42 14.24  & +06 27 52.4   & $18.9{4}\pm0.02$ & ${19.41\pm0.24}$ & ${-21.52\pm0.24}$    & $0.359$     & $0.023$     & photo \\
RGZ J103435.8+25181{7}   & 10 34 35.81  & +25 18 17.9   & $19.{8}6\pm{0.04}$ & ${19.51\pm0.06}$ & ${-22.21\pm0.06}$  & $0.39481$ & $0.00002$     & spectro \\
RGZ J105521.2+372652     & 10 55 21.24  & +37 26 52.6   & $18.{78\pm0.01}^{\$}$ & ${18.32\pm0.02}$ & ${-24.34\pm0.02}$& $0.58858$   & $0.00006$   & spectro \\
RGZ J105838.{6}+244535   & 10 58 38.66  & +24 45 35.1   & $17.9{4\pm0.01}$ & ${18.05\pm0.02}$ & ${-21.78\pm0.04}$    & $0.201$     & $0.016$     & photo \\
RGZ J120343.7+23430{4}   & 12 03 43.71  & +23 43 04.7   & $16.{66}\pm0.01$ & ${16.70\pm0.02}$ & ${-22.88\pm0.02}$    & $0.1767$    & $0.00005$   & spectro \\
RGZ J122425.8+02031{0}   & 12 24 25.84  & +02 03 10.7   & $18.{59}\pm0.02$ & ${18.66\pm0.05}$ & ${-22.92\pm0.05}$    & $0.45157$   & $0.00007$   & spectro \\
RGZ J122653.9+04291{8}   & 12 26 53.91  & +04 29 18.9   & $19.{47}\pm0.02$ & ${19.32\pm0.04}$ & ${-22.65\pm0.04}$    & $0.51743$   & $0.00002$   & spectro \\
RGZ J123300.{2}+06032{5} & 12 33 00.30  & +06 03 26.1   & ${19.16}\pm0.02$ & ${19.08\pm0.04}$ & ${-21.80\pm0.11}$    & $0.269$     & $0.058$     & photo \\
RGZ J123414.{7}+22224{8} & 12 34 14.73  & +22 22 47.9   & ${23.13\pm0.56}$ & $- ^{\#}$                & $- ^{\#}$      & $0.881$     & $0.052$     & photo \\
RGZ J131414.{1}+02040{4} & 13 14 14.19  & +02 04 04.3   & $21.{45\pm0.13}$ & ${21.52\pm0.41}$ & ${-22.22\pm0.43^\diamond}$ & $0.982$  & $0.086$     & photo \\
RGZ J144300.1+144042     & 14 43 00.11  & +14 40 42.1   & $20.2{6\pm0.04}$ & ${20.90\pm0.13}$ & ${-20.58\pm0.15}$    & $0.425$      & $0.041$     & photo \\
RGZ J144921.5+50194{5}   & 14 49 21.53  & +50 19 45.3   & $19.2{0}\pm0.02$ & ${19.30\pm0.04}$ & ${-21.71\pm0.11}$    & $0.329$      & $0.061$     & photo \\
RGZ J150407.5+574918     & 15 04 08.08  & +57 49 22.5   & ${20.28\pm0.03}$ & ${20.04\pm0.07}$ & ${-21.98\pm0.11}$    & $0.465$      & $0.051$     & photo \\
RGZ J150455.{5}+56492{0} & 15 04 55.56  & +56 49 20.3   & ${19.92\pm0.03}$ & ${20.36\pm0.18}$ & ${-20.57\pm0.18}$    & $0.35871$    & $0.00004$   & spectro \\
RGZ J151136.9+335501     & 15 11 36.93  & +33 55 01.1   & $20.{36\pm0.06}$ & ${20.18\pm0.08}$ & ${-23.01\pm0.08^\diamond}$ & $0.62341$& $0.00020$   & spectro \\
RGZ J152737.1+18225{0}   & 15 27 37.11  & +18 22 51.3   & $20.{50\pm0.10}$ & ${21.11\pm0.27}$ & ${-20.14\pm0.29}$    & $0.445$      & $0.074$     & photo \\
RGZ J153421.4+333436     & 15 34 21.43  & +33 34 35.9   & $18.{15}\pm0.01$ & ${18.25\pm0.02}$ & ${-21.70\pm0.04}$    & $0.210$      & $0.015$     & photo \\
RGZ J170002.6+270549     & 17 00 03.15  & +27 05 50.6   & ${20.69\pm0.12}$ & ${20.64\pm0.19}$ & ${-22.24\pm0.19^\diamond}$ & $0.719$  & $0.032$     & photo \\
\hline
\end{tabular}
\flushleft
{\bf Notes:} $\dagger$ \cite{1994AJ....107.1245S}. $\ddagger$ Average of the SDSS DR13 (KD-tree method), SDSS DR10 (RF method), \cite{2014A&A...568A.126B} and \cite{2016ApJS..225....5B} estimates. $*$ Calculated with SDSS $g$-band magnitude that has been verified with the Pan-STARRS measurement \cite[][]{2016arXiv161205243F,2016ApJ...822...66F}. $\$ $ Quasar, psf magnitude used. \# Unreliable $g$-band magnitude. $\diamond$ Lower limit, correct value of k-correction unavailable due to high redshift of the galaxy.
\end{minipage}


\footnotesize
\caption{Mid-IR hosts and radio luminosity densities of candidate RGZ HyMoRS. $S_{\rm GLEAM/NVSS/GB6}$ { are} integrated flux densities as measured in the GLEAM { (200~MHz)}, NVSS { (1.4~GHz)} or GB{ 6 (4.85~GHz)} surveys. $L_{\rm NVSS}$ is the rest frame total luminosity density of the source at 1.4~GHz, and $\alpha$ is the spectral index (simple power-law) calculated from the GLEAM, NVSS and/or GB6 data. \vspace{-4mm}}
\label{tab:wise-radio}
\begin{center}
\begin{tabular}{llccccccc}
\hline
\hline
HyMoRS name   & All{\it WISE} host name & $S_{\rm GLEAM}$ & $S_{\rm NVSS}$ & $S_{\rm GB6}$ & \multicolumn{1}{c}{$L_{\rm NVSS}$}& $\alpha$ \\
              &                         & (mJy)        & (mJy)        & (mJy)      & ($\times 10^{24}$\,W\,Hz$^{-1}$\,sr$^{-1}$)  &          \\
\hline
RGZ J023832.{6}+023349          & J023832.67+023349.0 & $1600\pm160$ & $438.1\pm11.9$ & $128\pm12$ &  ${4.85\pm0.13}$ & $0.79\pm0.06$ \\
RGZ J072406.{7}+38034{8}        & J072406.79+380348.6 & --           & $323.5\pm10.5$ & $77\pm7$   &  $4.60\pm0.15$ & $1.16\pm0.08$ \\
RGZ J082234.{0}+53111{8}        & J082234.06+531118.9 & --           &  $75.7\pm 2.6$ & $35\pm5$   &  $0.34\pm0.01$ & $0.62\pm0.12$ \\
RGZ J083352.2+04582{2}$^\ddagger$ & J083352.24+045822.6 &  $532\pm 43$ & $149.9\pm 5.1$ & $39\pm6$   &  ${2.02\pm0.28}$ & $0.76\pm0.07$ \\
RGZ J084738.{0}+183156          & J084738.07+183156.2 &  $178\pm 23$ &  $24.5\pm 1.2$ & --         &   --           & $1.02\pm0.07$ \\
RGZ J085926.7+292738            & J085926.73+292738.1 & $1847\pm240$ & $547.1\pm15.7$ & $176\pm16$ & $11.22\pm0.42$ & $0.75\pm0.06$ \\
RGZ J091408.0+52294{ 8}         & J091408.03+522948.7 & --           &  $30.9\pm 1.5$ & --         &  $4.45\pm0.27$ & --            \\
RGZ J094214.2+062752$^\ddagger$   & J094214.21+062751.9 &  $265\pm 21$ &  $81.0\pm 2.8$ & --         &  $3.34\pm0.14$ & $0.61\pm0.04$ \\
RGZ J103435.8+25181{7}          & J103435.81+251817.8 &  $263\pm 34$ &  $62.0\pm 2.5$ & --         &  $3.09\pm0.12$ & $0.74\pm0.07$ \\
RGZ J105521.2+372652            & J105521.24+372652.4 & --           &  $92.5\pm 2.5$ &  $30\pm 4$ & $11.44\pm0.31$ & $0.91\pm0.11$ \\
RGZ J105838.{6}+244535          & J105838.67+244535.0 &  $886\pm109$ & $158.5\pm 4.9$ & -- ($\diamond$) &  ${1.58}\pm0.06$ & $0.88\pm0.07$ \\
RGZ J120343.7+23430{4}          & J120343.72+234304.6 & $1586\pm206$ & $430.3\pm10.4$ &  $81\pm 8$ &  $3.18\pm0.08$ & $0.95\pm0.13$ \\
RGZ J122425.8+020311            & J122425.83+020310.6 &  $417\pm 33$ & $69.8\pm 2.1$  & --         &  $4.52\pm0.14$ & $0.92\pm0.04$ \\
RGZ J122653.9+04191{8}          & J122653.90+042918.9 & $2717\pm217$ & $688.9\pm21.7$ & $245\pm23$ & $66.52\pm2.10$ & $0.75\pm0.02$ \\
RGZ J123300.{2}+06032{5}        & J123300.28+060325.6 &  $245\pm 20$ &  $42.6\pm 1.7$ & --         &  $0.82\pm0.06$ & $0.90\pm0.05$ \\
RGZ J123414.{7}+22224{8}        & J123414.79+222248.9 &  $349\pm 45$ &  $63.4\pm 1.9$ &  $22\pm 4$ & $21.82\pm1.23$ & $0.87\pm0.01$ \\
RGZ J131414.{1}+02040{4}        & J131414.19+020404.6 &  $216\pm 17$ &  $47.8\pm 1.9$ & --         & $23.00\pm2.01$ & $0.78\pm0.05$ \\
RGZ J144300.1+144042            & J144300.11+144042.2 &  $442\pm 35$ &  $91.2\pm 2.6$ &  $35\pm 5$ &  $5.33\pm0.24$ & $0.80\pm0.01$ \\
RGZ J144921.5+50194{5}          & J144921.52+501945.7 & --           & $137.7\pm 5.0$ &  $44\pm 5$ &  $4.21\pm0.32$ & $0.92\pm0.10$ \\
RGZ J150407.5+574918$^\ddagger$   & J150407.50+574918.0 & --           &  $97.5\pm 3.5$ & --         &  $7.23\pm0.46$ & --            \\
RGZ J150455.{5}+56492{0}        & J150455.56+564920.5 & --           & $118.6\pm 4.5$ &  $61\pm 6$ &  $4.99\pm0.19$ & $0.54\pm0.08$ \\
RGZ J151136.9+335501            & J151136.94+335501.1 & --           &  $65.6\pm 2.3$ &  $21\pm 4$ &  $9.29\pm0.33$ & $0.92\pm0.16$ \\
RGZ J152737.1+18225{0}          & J152737.11+182250.7 &  $679\pm 54$ & $143.5\pm 4.9$ &  $72\pm 7$ &  $9.66\pm0.81$ & $0.72\pm0.05$ \\
RGZ J153421.4+333436            & J153421.42+333436.0 & --           & $239.8\pm 8.2$ &  $85\pm 8$ &  $2.67\pm0.10$ & $0.83\pm0.08$ \\
RGZ J170002.6+270549$^\ddagger$   & --                  & --           &  $54.2\pm 2.0$ & --         & $11.96\pm0.57$ & --            \\
\hline
\end{tabular}
\end{center}\vspace{-3mm}
{\bf Notes:}  $\ddagger$ Host uncertain or confused in the {\it WISE} image. $\diamond$ Only one lobe detected.\\
\end{table*}


\subsection{Radio Galaxy Zoo Catalog}

The first data release of the Radio Galaxy Zoo project (hereafter RGZ DR1) is based on the project's first 2.5 years of operation (Wong et al. {\it in prep.}). The RGZ DR1 catalog consists of over 74,000 radio source components from the FIRST survey, with weighted consensus level of 65\% or greater, where the consensus means here the level of agreement on the chosen mid-IR host and radio components of a subject being classified. A weighted consensus level of 65\% results in a classification that is reliable at the 80\%, or greater level (Wong et al. {\it in prep.}). Radio source classifications that are derived from the RadioTalk forum (such as those in this paper) are more likely to have greater reliability, since they stem from discussions between the citizen scientists and the science team. In this paper we investigate all entries from the RGZ project pipeline of our {candidate HyMoRS}, including those entries that fall below the RGZ DR1 catalog consensus lower limit of 65\%. For more details on the classification see \cite{2015MNRAS.453.2326B} and Wong et al. ({\it in prep.}).

\subsection{Multi-wavelength cross-matching}

We repeat the work of the citizen scientists and manually cross-match each selected RGZ radio source with the mid-IR {\it WISE} ($3.4\mu$m band) and optical $r$-band ($623$nm) Sloan Digital Sky Survey Data Release 13 \cite[SDSS DR13;][]{2016arXiv160802013S} to  verify the radio morphology of the sources, and to obtain redshift estimates. The manual selection of the mid-IR hosts allows us to cross-check the accuracy of the RGZ DR1 catalog specifically for future automated identification of HyMoRS candidates. 

In addition, we cross-match our selected sources with radio surveys at other radio frequencies and angular resolution lower than that of the FIRST survey: the Galactic and Extragalactic All-Sky Murchison Widefield Array (MWA) survey \cite[GLEAM;][]{2017MNRAS.464.1146H} at 200~MHz, the NRAO Very Large Array (VLA) Sky Survey \cite[NVSS;][]{1998AJ....115.1693C} at 1.4~GHz, and the Green Bank 6cm survey \cite[GB6;][]{1996ApJS..103..427G} at 4.85~GHz. This procedure is to obtain total flux density measurements of our sources at 1.4~GHz, and an overall radio spectral index ($\alpha$, where the flux density $S$ at a frequency $\nu$ is $S_{\nu}\propto \nu^{-\alpha}$) for the radio luminosity density redshift correction.


\begin{table*}
\footnotesize
\caption{Radio properties of candidate HyMoRS. $f_{\rm FR}$ is an index that quantifies FR morphology based on the position of the brightest feature in the source's lobe, calculated between the position of the optical host and farthest extent of the lobe and separately for each lobe of each radio galaxy (see \S\ref{sec:radio} for details). Values quoted in brackets are for cases where the radio core dominates the lobe flux density. $\theta_{\rm FRI/FRII}$ is {the} projected angular extent of either FRI or FRII lobe, measured between the position of the host and outermost $3\sigma$ contours. $\theta_{\rm total}$ is {the} projected total angular extent of the radio source measured between outermost $3\sigma$ contours with an accuracy of 5 arcsec. The FIRST survey images are used for the size measurements unless stated otherwise. $D_{\rm total}$ is the total linear size of the source.  \vspace{-4mm}}
\label{tab:radio-ratio}
\begin{center}
\begin{tabular}{lllcccr}
\hline
\hline
HyMoRS name      & $f_{\rm FR}$ / FRI side         & $f_{\rm FR}$ / FRII side & $\theta_{\rm FRI}$ & $\theta_{\rm FRII}$ & $\theta_{\rm total}$   & $D_{\rm total}$ \\
                 &                               &                        & (arcsec)         & (arcsec)         &  (arcsec)            & (kpc)\\
\hline
RGZ J023832.{6}+023349   & $1.39\pm0.07^{\dagger}$         & $2.35\pm0.04$         & 355$^{\dagger}$          & 170$^{\dagger}$ & 530$^{\dagger}$        & ${1860\pm20}$\\
RGZ J072406.{7}+38034{8} & $2.02\pm0.07$ ($0.50\pm0.01$) & $2.07\pm0.07$         & 61                       & 69         & 130                  & $510\pm20$ \\
RGZ J082234.{0}+53111{8} & $0.96\pm0.24$ ($0.50\pm0.01$) & $2.18\pm0.10$         & 25                       & 47         &  70                  & $175\pm15$ \\
RGZ J083352.2+04582{2}   & $1.33\pm0.07$                 & $2.26\pm0.10$         & 65                       & 47         & 110                  & ${410\pm65}$ \\
RGZ J084738.{0}+183156   & $1.79\pm0.17$                 & $2.20\pm0.12$         & 27                       & 38         &  65                  & --         \\ 
RGZ J085926.7+292738     & $1.58\pm0.09$ ($0.50\pm0.01$) & $2.23\pm0.08$         & 49                       & 56         & 100                  & $470\pm25$ \\
RGZ J091408.0+52294{8}   & $1.43\pm0.16$                 & $2.21\pm0.12$         & 29                       & 40         &  70                  & $480\pm40$ \\
RGZ J094214.2+06275{1}   & $1.02\pm0.13$                 & $1.61\pm0.27$         & 43                       & 18         &  60                  & $310\pm25$ \\
RGZ J103435.8+25181{7}   & $1.17\pm0.18$ ($0.50\pm0.01$) & $2.21\pm0.12$ ($0.60\pm0.43$) & 29               & 38         &  70                  & $380\pm25$ \\
RGZ J105521.2+372652     & $1.45\pm0.12$                 & $2.20\pm0.09$         & 40                       & 51         &  85                  & $575\pm35$ \\
RGZ J105838.{6}+244535   & $1.50\pm0.05^{\dagger}$         & $2.18\pm0.05^{\dagger}$  & 340$^{\dagger}$         & 330$^{\dagger}$ & 670$^{\dagger}$        & $2280\pm55$\\
RGZ J120343.7+23430{4}   & $1.17\pm0.04$                 & $2.27\pm0.03$          & 110                     & 145        & 255                  & $785\pm15$ \\
RGZ J122425.8+02031{0}   & $1.78\pm0.10$                 & $2.27\pm0.09$          & 47                      & 49         &  95                  & $560\pm30$ \\
RGZ J122653.9+04191{8}   & $1.55\pm0.13$                 & $2.13\pm0.09$          & 36                      & 51         &  85                  & $540\pm30$ \\
RGZ J123300.{2}+06032{5} & $1.07\pm0.20$                 & $2.15\pm0.14$          & 27                      & 33         &  60                  & $255\pm30$ \\
RGZ J123414.{7}+22224{8} & $1.57\pm0.09$                 & $2.33\pm0.07$          & 52                      & 67         & 115                  & $910\pm55$ \\
RGZ J131414.{1}+02040{4} & $1.68\pm0.14$ ($0.50\pm0.10$) & $2.18\pm0.13$          & 36                      & 36         &  65                  & $530\pm55$ \\
RGZ J144300.1+144042     & $1.06\pm0.11$                 & $2.24\pm0.06$          & 47                      & 72         & 120                  & $685\pm40$ \\
RGZ J144921.5+50194{5}   & $1.26\pm0.12$ ($0.60\pm0.43$) & $2.21\pm0.12$ ($0.50\pm0.01$) & 40               & 40         &  75                  & $365\pm35$ \\
RGZ J150407.5+574918     & $1.30\pm0.10$                 & $2.06\pm0.13$          & 47                      & 34         &  80                  & $480\pm40$ \\
RGZ J150455.{5}+56492{0} & $0.93\pm0.14$                 & $2.35\pm0.05$          & 43                      & 97         & 140                  & $720\pm25$ \\
RGZ J151136.9+335501     & $0.76\pm0.18$                 & $1.83\pm0.13$          & 43                      & 34         &  75                  & $520\pm35$ \\
RGZ J152737.1+18225{0}   & $1.35\pm0.08$ ($0.50\pm0.10$) & $2.07\pm0.10$          & 61                      & 43         &  75                  & $440\pm45$ \\
RGZ J153421.4+333436     & $0.91\pm0.11$                 & $2.18\pm0.10$          & 54                      & 47         & 100                  & $350\pm20$ \\
RGZ J170002.6+270549     & $1.17\pm0.15$ ($0.61\pm0.44$) & $1.64\pm0.17$          & 33                      & 27         &  60                  & $440\pm40$ \\
\hline
\end{tabular}
\end{center}\vspace{-3mm}
{\bf Notes:}  $\dagger$ NVSS survey used. \\
\end{table*}


\begin{table*}
\caption{Candidate HyMoRS: RGZ DR1 Catalog entries.}
\label{tab:rgzcat}
\begin{tabular}{lcccc p{4.8cm}}
\hline
\hline
HyMoRS name  & No. of catalog  & Correct mid-IR   & All radio components & Consensus & Comments\\
             & entries         & host identified? & included?            &           & \\
\hline
RGZ J072406.{7}+38034{8}   & 1   & yes  & no   & 57\%  & One radio component missing (outside RGZ cut-out).\\
RGZ J082231.1+531119       & 1   & yes  & yes  & 62\% \\
RGZ J085926.7+292738       & 1   & yes  & yes  & 100\% \\
RGZ J091408.0+52294{8}     & 1   & yes  & yes  & 84\% \\
RGZ J094214.2+06275{1}     & 1   & yes  & yes  & 86\% \\
RGZ J105521.2+372652       & 1   & yes  & yes  & 62\% \\ 
RGZ J120343.7+234205       & 1   & no   & no   & 55\%  & SE lobe only in the RGZ cut-out. Correctly assigned no IR host, but wrong host selection of the overall radio galaxy (see \S\ref{sec:hymors-cat}).\\
RGZ J123414.{7}+22224{8}   & 1   & yes  & yes  & 52\% \\
RGZ J131414.{1}+02040{4}   & 1   & yes  & yes  & 75\% \\
RGZ J144300.1+144042 & 3   & 	  &      &       & Both lobes seen in the RGZ cut-outs.\\
                     & --a & no   & no   & 71\%  & Only S lobe selected. Mid-IR object superposed within the lobe extent selected as host.\\
                     & --b & yes  & yes  & 43\%\\ 
                     & --c & no   & yes  & 70\%  & All radio components correctly identified as part of the radio galaxy, but radio galaxy incorrectly assigned no IR host.\\
RGZ J150407.5+574918 & 1   & yes  & yes  & 81\%\\	
RGZ J151136.9+335501 & 2   & 	  &      & \\
                     & --a & yes  & yes  & 86\%\\
                     & --b & no   & yes  & 78\%  & Incorrect host selected. \\
RGZ J152737.1+18225{0} & 2   & 	  &      & \\
                     & --a & yes  & yes  & 69\%\\
                     & --b & yes  & yes  & 49\%\\
RGZ J153421.5+333436 & 1   & yes  & yes  & 84\%\\
RGZ J170002.6+270549 & 1   & yes  & yes  & 80\%\\
\hline\\
\end{tabular}
\end{table*}


\section{Results}
\label{sec:results}

We select 25 candidate hybrid morphology radio galaxies. Our final sample is presented in Figures~ \ref{rys:hymors-large} and \ref{rys:hymors} where we plot radio contours from the FIRST survey over the {\it WISE} $3.4\mu$m-band images. In Tables~\ref{tab:optical} and \ref{tab:wise-radio} we list mid-IR and optical hosts of the radio galaxies, and provide radio, optical and redshift information. In this section we provide brief notes on our candidates (\S\ref{sec:hymors-notes}), present more detailed results on radio (\S\ref{sec:radio}) and optical (\S\ref{sec:optical}) properties of the candidates, and compare our results to the RGZ DR1 catalog (\S\ref{sec:hymors-cat}).

\subsection{Notes on candidate HyMoRS}
\label{sec:hymors-notes}

\begin{figure*}
\includegraphics[width=100mm]{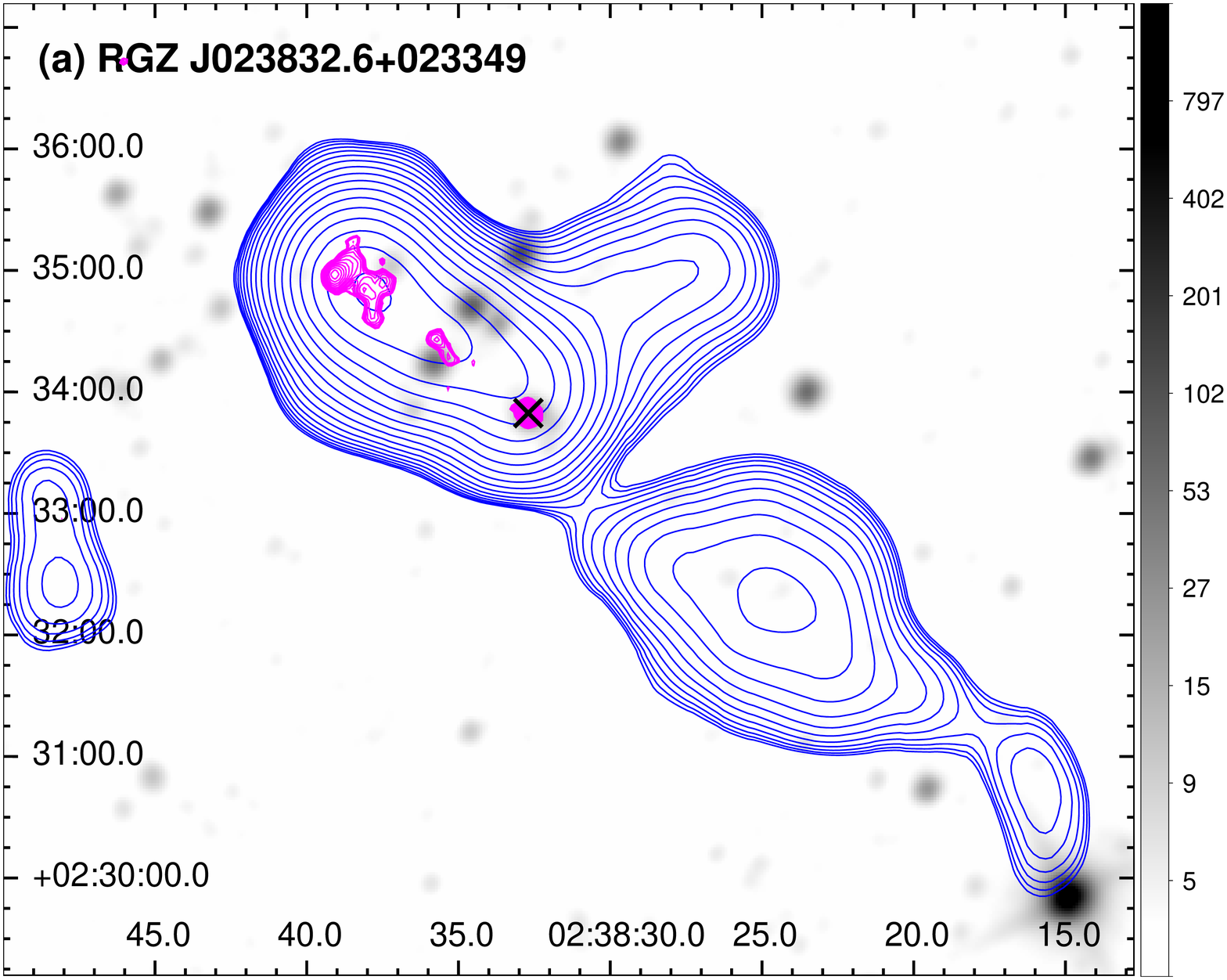}
\includegraphics[width=76mm]{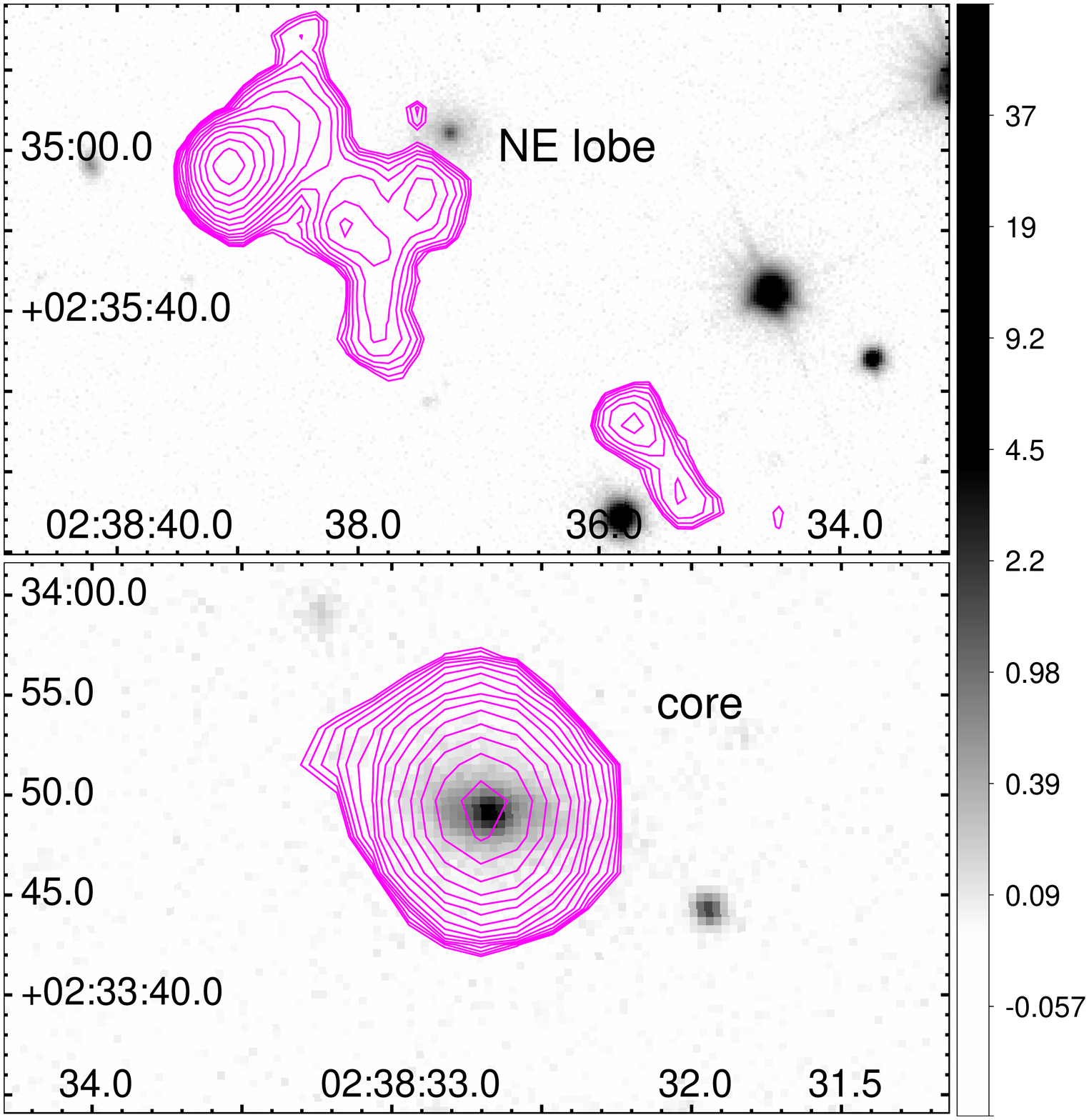}\\

\includegraphics[width=100mm]{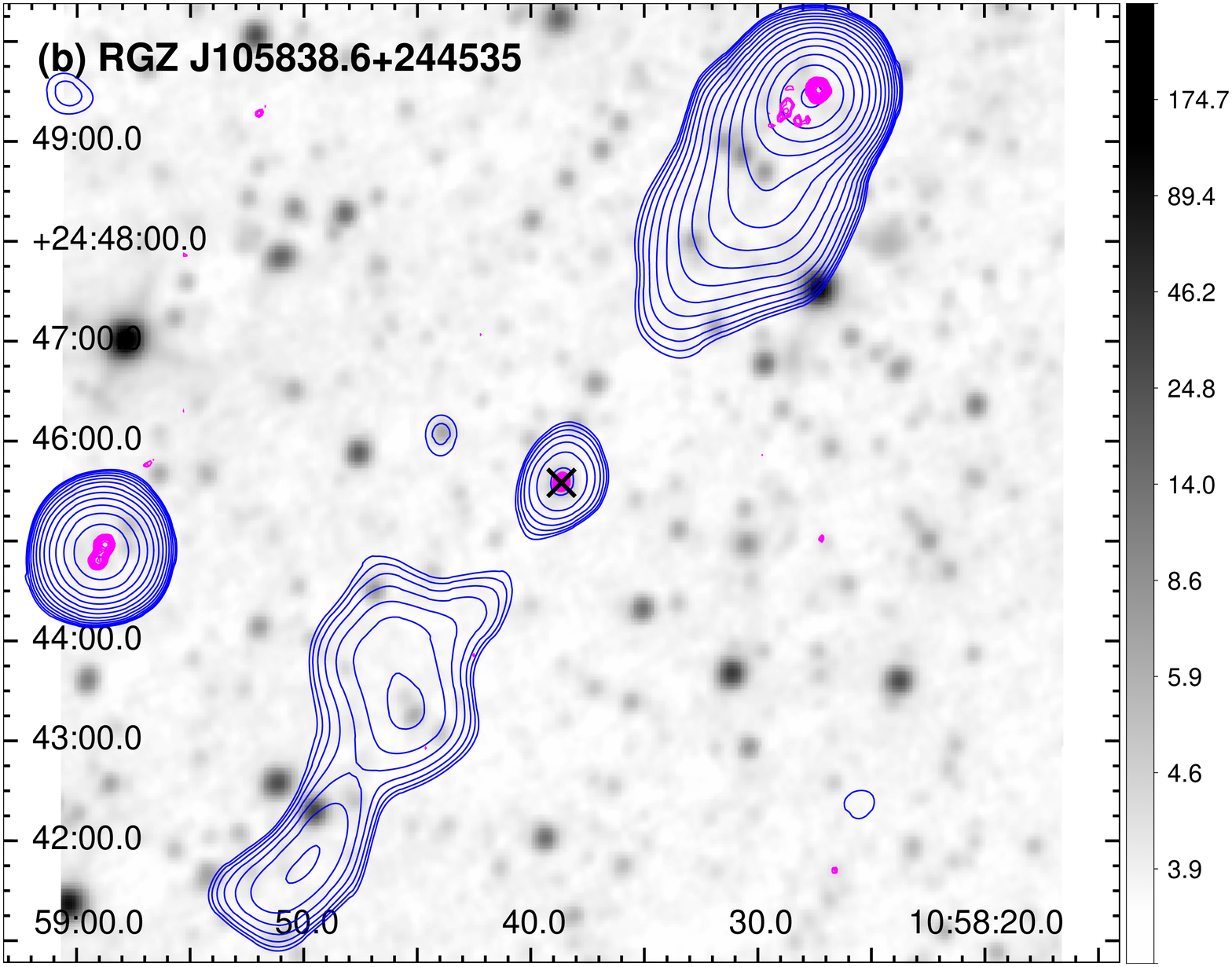}
\includegraphics[width=76mm]{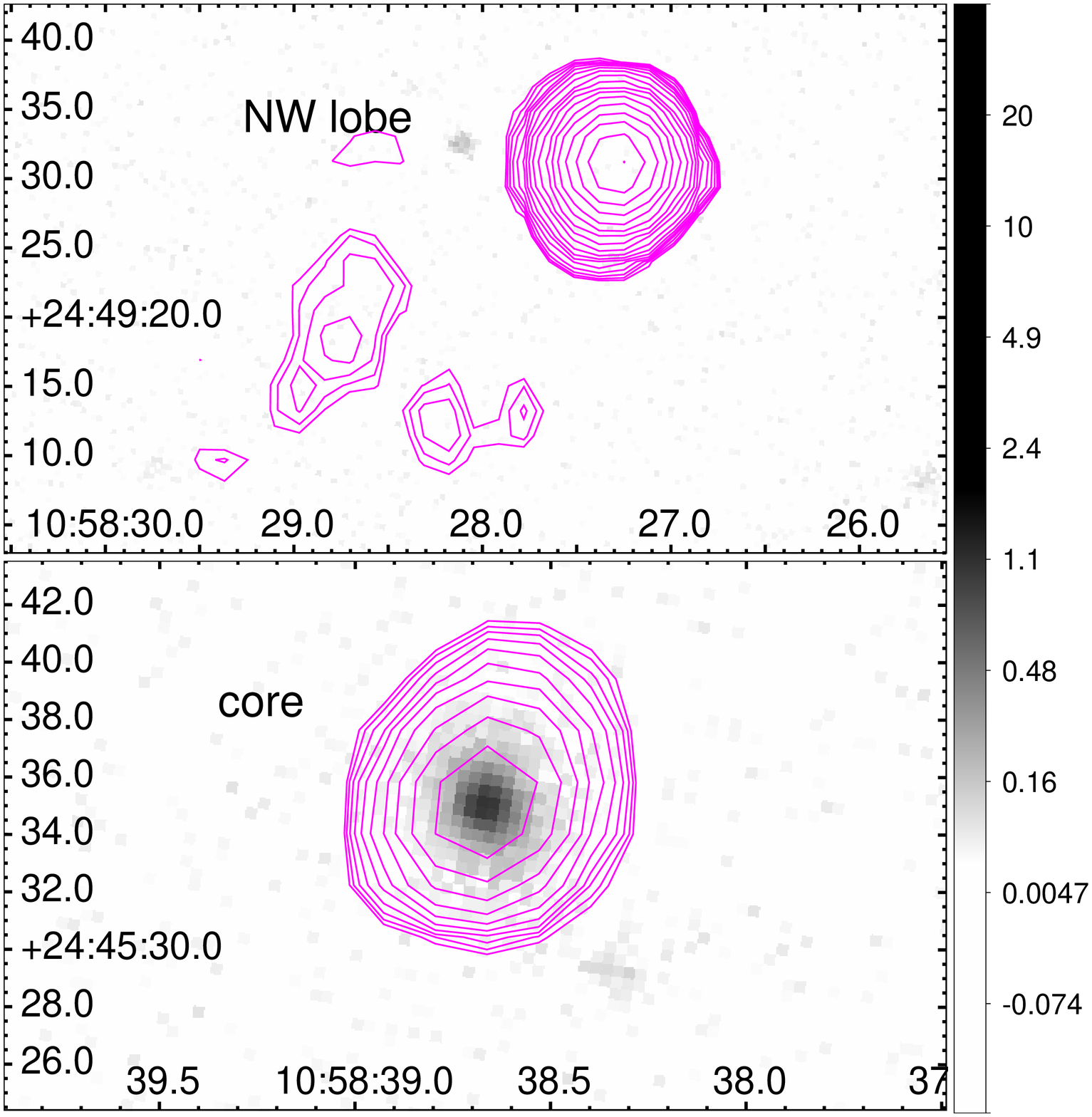}\\
\caption{{\it Left:} {\it WISE} $3.4\mu$m images with overlaid FIRST (magenta) and NVSS (blue) of two giant candidate HyMoRS for which inspection of the NVSS image was necessary. The contours are drawn at levels $3\sigma+\sigma\times2^{n/2}$ for $n\geq0$, and where $\sigma=0.15$~mJy~beam$^{-1}$ (FIRST) and $\sigma=${0.45}~mJy~beam$^{-1}$ (NVSS). {\it WISE}: The colourbars are in log scale and are in the raw intensity units of DN/pixel. {H}osts of the radio galaxies are marked with crosses. {\it Right:} Zoom onto components detected in FIRST overlaid on SDSS DR13 $r'$ band images. The FIRST contours are at the same level as in the left panel images. The colourbars are in log scale and are in the SDSS units of nanomaggies\footnote{\tt http://www.sdss.org/dr13/algorithms/magnitudes/ \\}.}
\label{rys:hymors-large}
\end{figure*}

1. {\bf RGZ J023832.6+023349} (Figure~\ref{rys:hymors-large}a): In the FIRST image the NE lobe displays a strong hotspot-like component at its far end. The SW lobe is not detected in the FIRST image, but faint diffuse emission is detected in the NVSS image with no compact components present. A compact radio core detected in the FIRST image is coincident with a mid-IR host (Table~\ref{tab:wise-radio}). This is a giant ($>1$~Mpc) radio galaxy, with a size of almost 2~Mpc \cite[{based on work in} ][]{2012sngi.confP...1A}, and one of the examples where the low angular resolution NVSS images are needed for the classification of the radio source. This source is a QSO at a spectroscopic redshift $0.209$ \cite[][]{1994AJ....107.1245S}.
 
2. {\bf RGZ J072406.7+380348} (Figure~\ref{rys:hymors}a): The SW lobe features a strong component, which can be considered a recessed hotspot, embedded in diffuse emission that links the hotspot to the radio core. The NE lobe has no dominant compact hotspot-like sources and is broken into separate components in the FIRST image. A compact radio core is detected. The host has targeted SDSS spectroscopic observations; located at $z=0.241$ it is classified as a QSO. 

3. {\bf RGZ J082231.1+531118} (Figure~\ref{rys:hymors}b): The NW lobe displays a brightness peak at its far end, which can be considered a hotspot-like component, with emission extending towards the radio core. The SW lobe is plume-like extending in the S and SE directions. A radio core is detected, but at {the} angular resolution of FIRST it is merged with the diffuse emission of the SW lobe.

4. {\bf RGZ J083352.2+045822} (Figure~\ref{rys:hymors}c): No radio core is detected, which makes the mid-IR host identification more difficult. In our interpretation the observed radio structure is {that} of a single radio galaxy with the mid-IR host {\it WISE} J083352.25+045822.7. The NE lobe hosts a strong hotspot-like component at its northernmost end. The SW lobe displays elongated, relaxed structure. We note that the structure could be also interpreted as coming from two separate radio galaxies, but in such a case the northern source would be composed of a single lobe with no distinguishable radio core. No additional low-level emission that could be associated to a potential southern lobe in such an alternative interpretation is detected in the NVSS image.

5. {\bf RGZ J084738.0+183158} (Figure~\ref{rys:hymors}d): The side NE from the radio core is a strong, compact component, which could be considered a hotspot. There is an optical object ({with} no mid-IR counterpart) in the vicinity that could be, in principle, associated with the NE component; however, the object seems to be detected in the SDSS images only, and it is classified within SDSS as a star. The SW lobe displays a tail-like, relaxed structure. A radio core is detected coincident with a faint mid-IR host. We found no optical or X-ray counterpart of the host in the publicly available surveys. However, there is a non-cataloged object 4~arcsec SE from the mid-IR host detected in the Digitized Sky Survey red image that potentially could be the optical counterpart. This radio galaxy is our weakest candidate HyMoRS, as there is a possibility the NE component is an unassociated infrared faint radio source \cite[IFRS;][]{2014MNRAS.439..545C}.

6. {\bf RGZ J085926.7+292738} (Figure~\ref{rys:hymors}e): The southern lobe features a strong, hotspot-like component at its far end, with diffuse emission extending between the strong component and the radio core. The northern lobe is devoid of any strong compact component, displaying only diffuse emission similar to the lobe emission of the southern lobe. A compact radio core is detected, embedded in the diffuse lobe emission.

7. {\bf RGZ J091408.0+522948} (Figure~\ref{rys:hymors}f): The eastern lobe is dominated by a hotspot-like component. The western lobe displays elongated plume-like emission with a surface brightness decreasing with distance from the core. A compact radio core is detected. 

8. {\bf RGZ J094214.2+062751} (Figure~\ref{rys:hymors}g): The emission of the SW lobe is uniform along its whole extent, displaying a confined structure. If a hotspot is present, it is merged with the lobe emission. The NE lobe consist{\bf s} of an elongated, tail-like diffuse emission with the brightness peak in the inner part of the lobe, and shows a relaxed structure. A radio core is not easily distinguished and, if present, it is merged with the southern lobe of the source. 

9. {\bf RGZ J103435.8+251817} (Figure~\ref{rys:hymors}h): The western lobe has a strong, hotspot-like component at its far end, with diffuse lobe emission pointing towards the radio core. The eastern lobe is devoid of any strong, compact components, and its diffuse emission seems to bend toward {the} south. A radio core is detected, but is merged with the eastern lobe. The host has targeted SDSS Baryon Oscillation Spectroscopic Survey \cite[SDSS BOSS; ][]{2016arXiv160802013S} observations; located at $z=0.3941$ it is classified as a luminous red galaxy. This object is a high excitation radio galaxy (HERG; see \S\ref{sec:optical}).

10. {\bf RGZ J105521.2+372652} (7C 1052+3742; Figure~\ref{rys:hymors}i): The northern lobe is dominated by a hotspot-like component. The southern lobe consists of an elongated diffuse emission of a relaxed structure. A compact radio core is detected. The host has targeted SDSS spectroscopic observations; located at $z=0.5886$ it is classified as a QSO.


\begin{figure*}
\includegraphics[width=60mm]{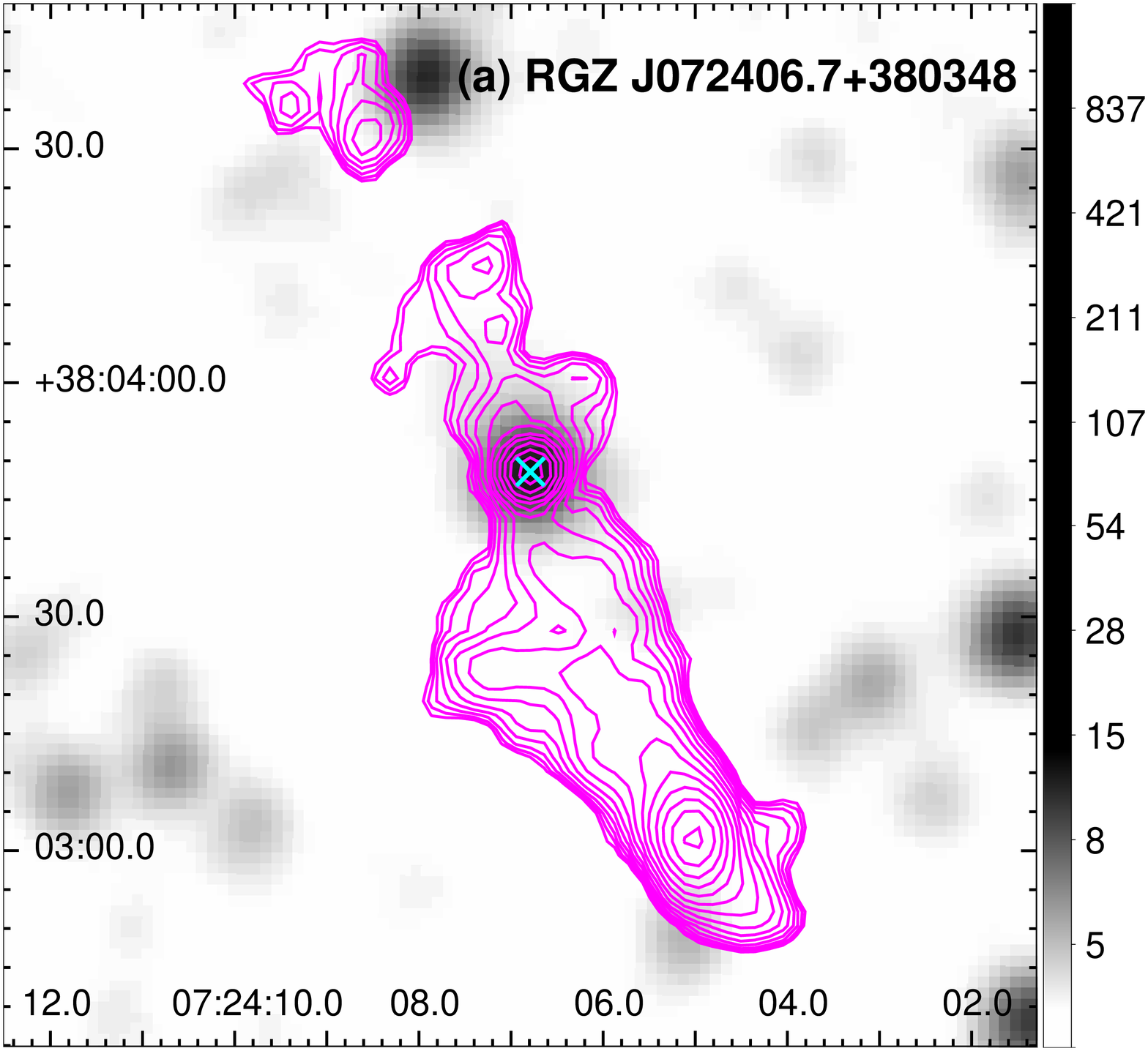}
\includegraphics[width=60mm]{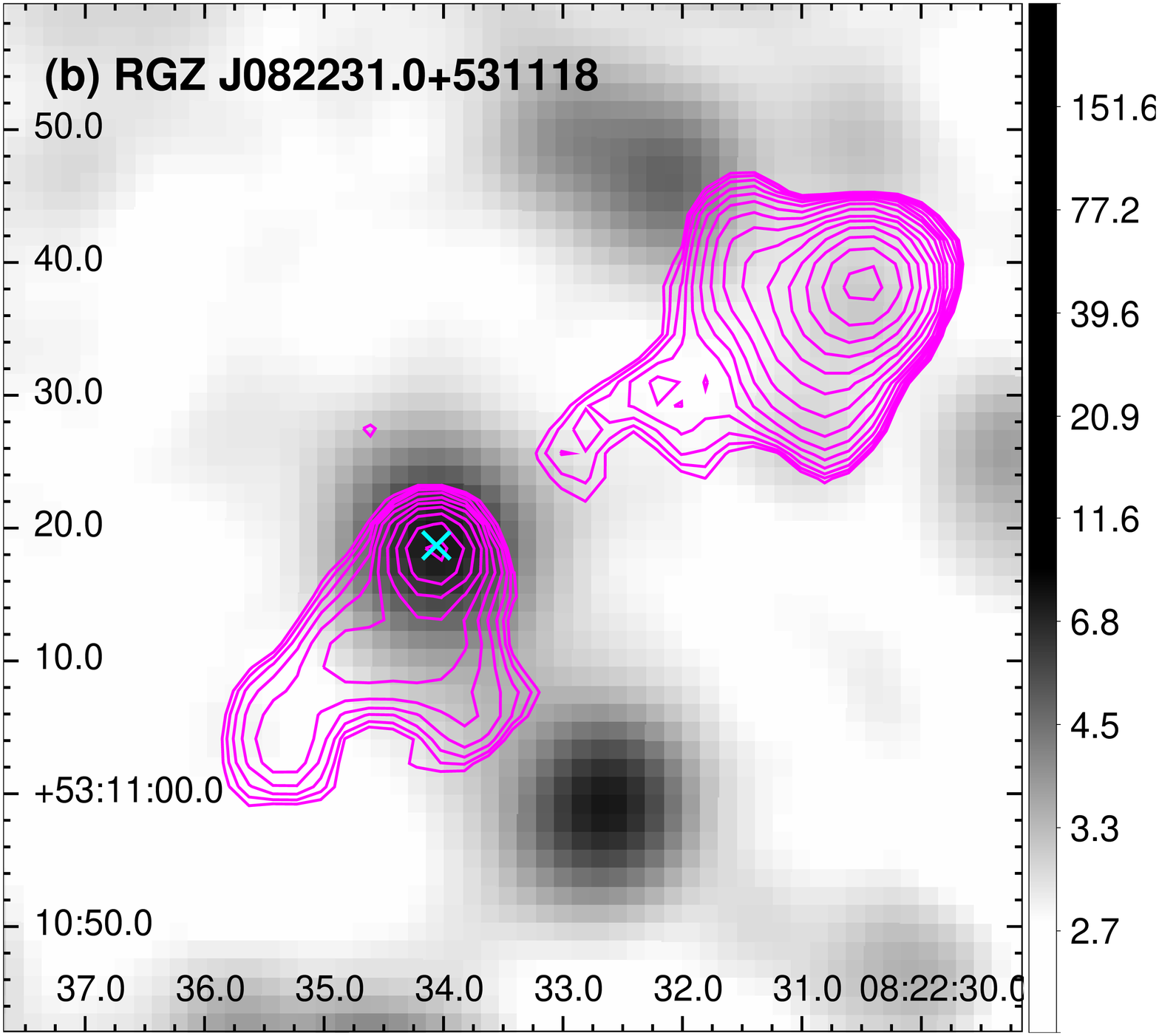}
\includegraphics[width=60mm]{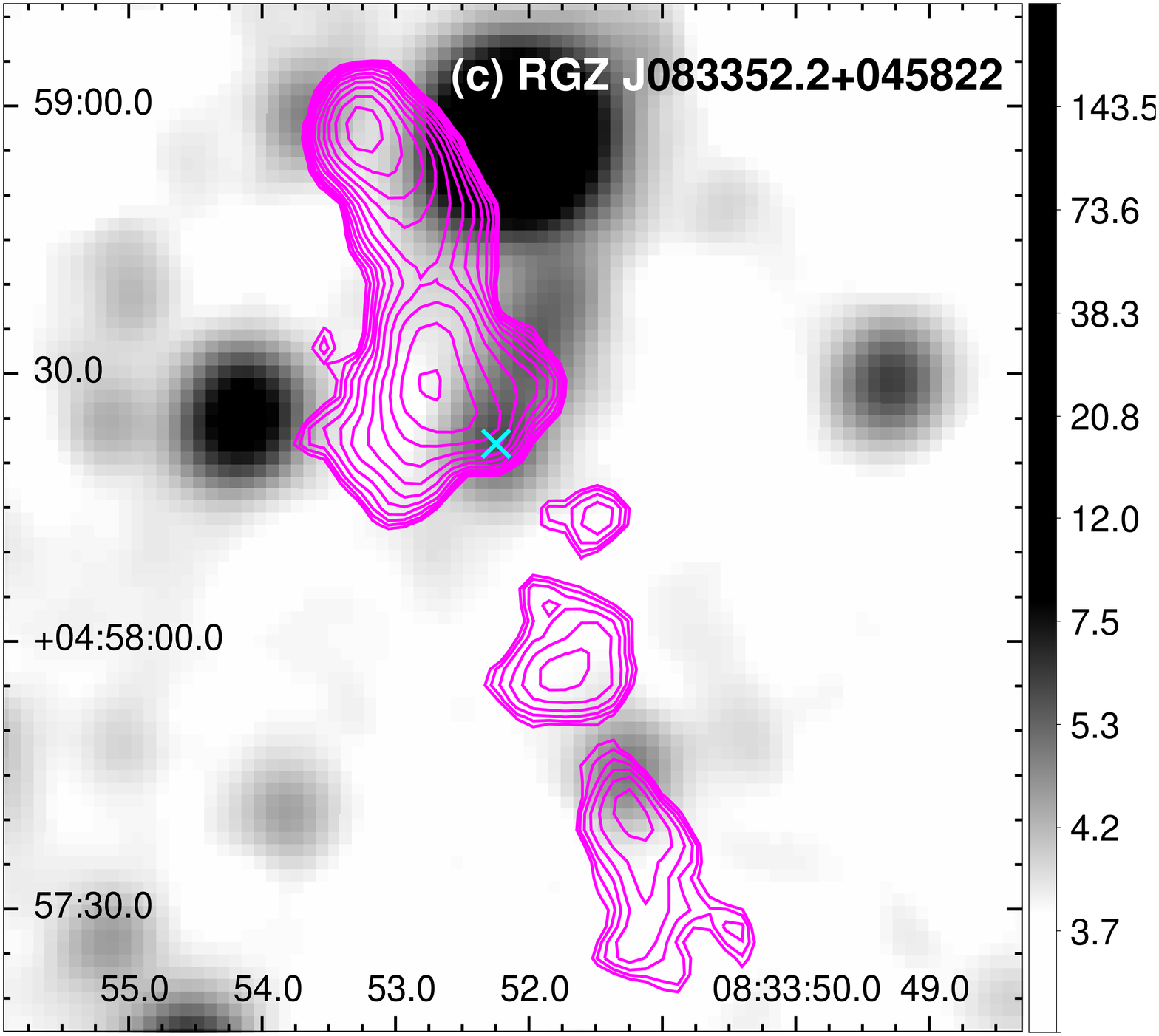}

\includegraphics[width=60mm]{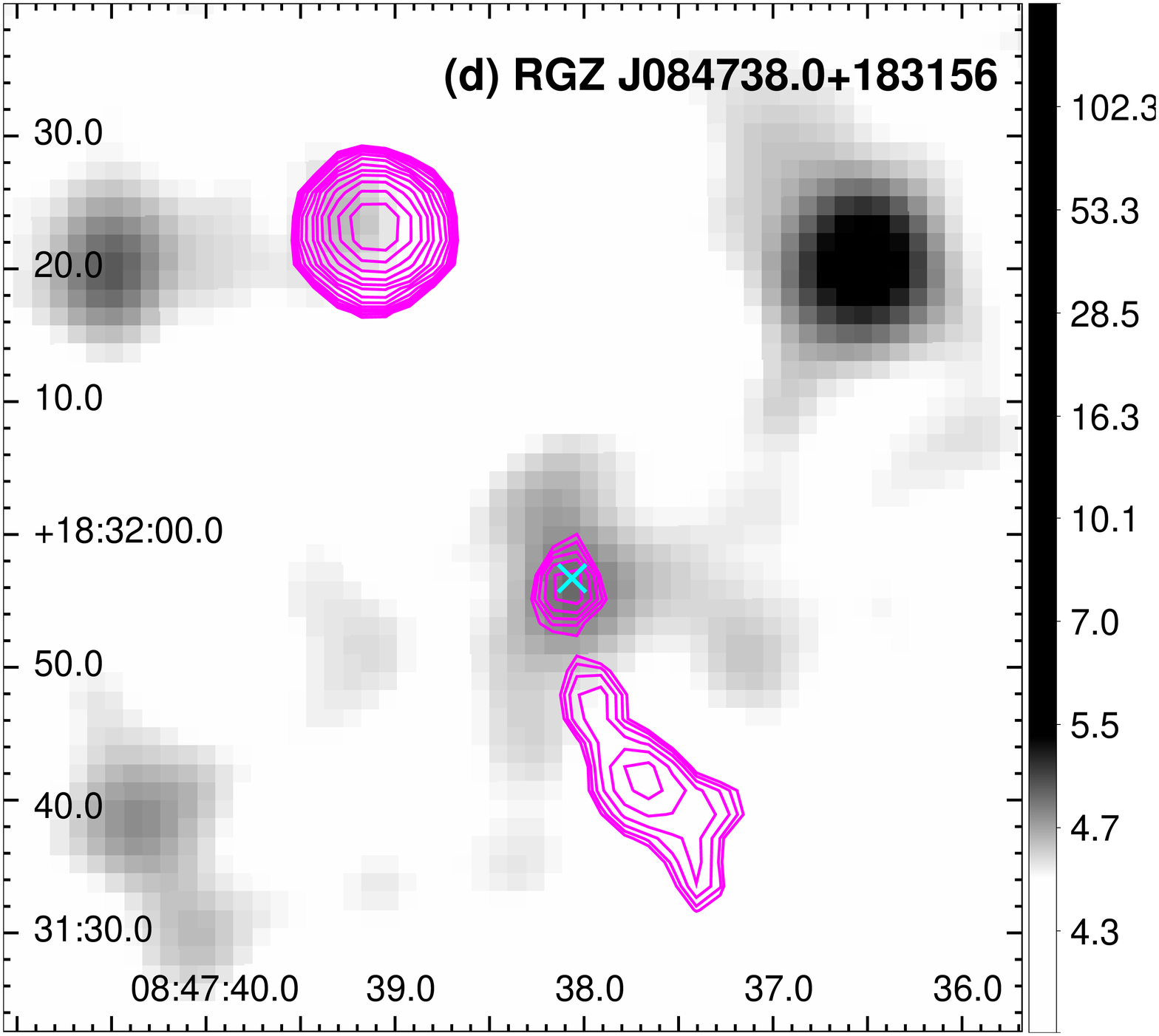}
\includegraphics[width=60mm]{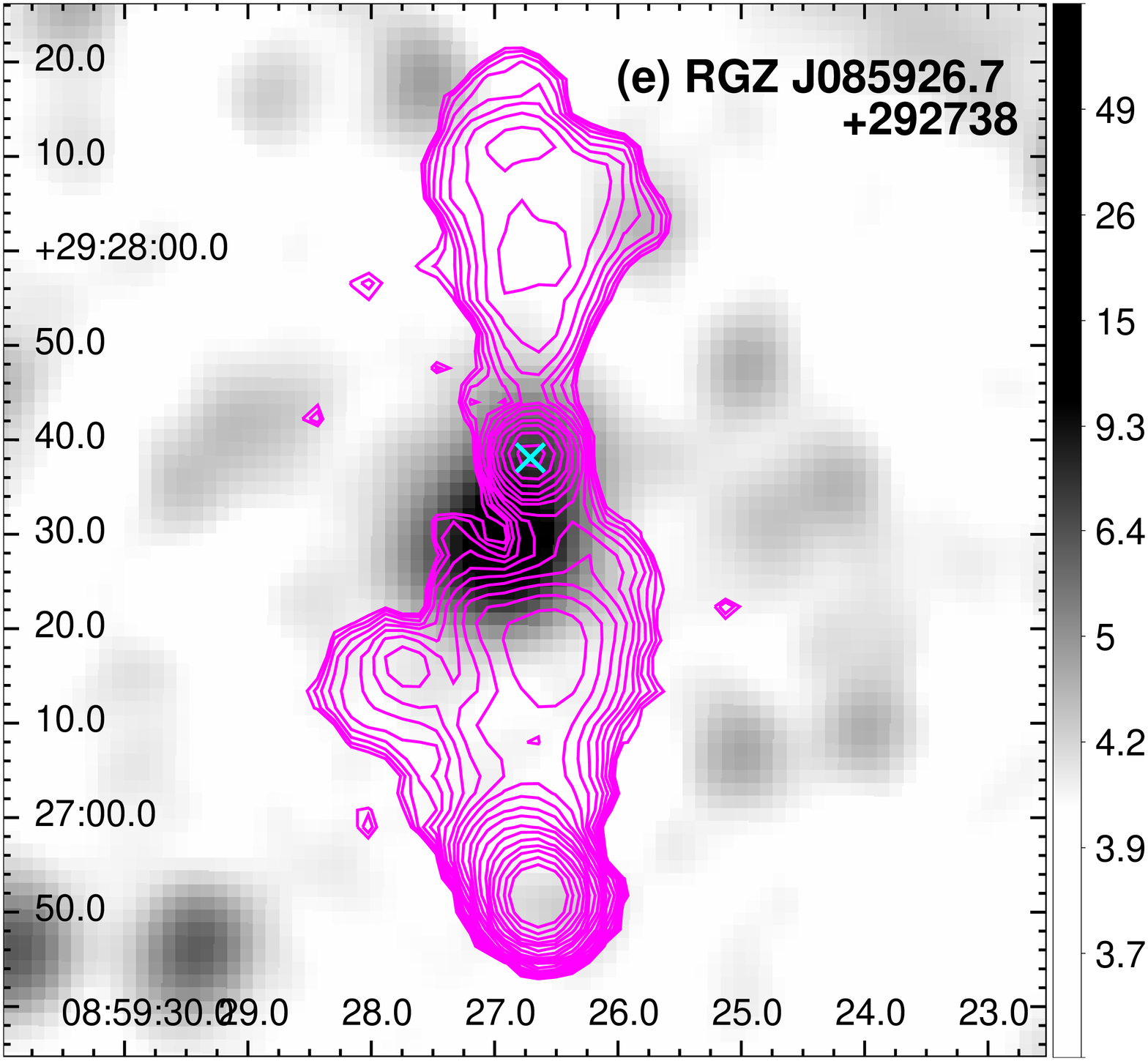}
\includegraphics[width=60mm]{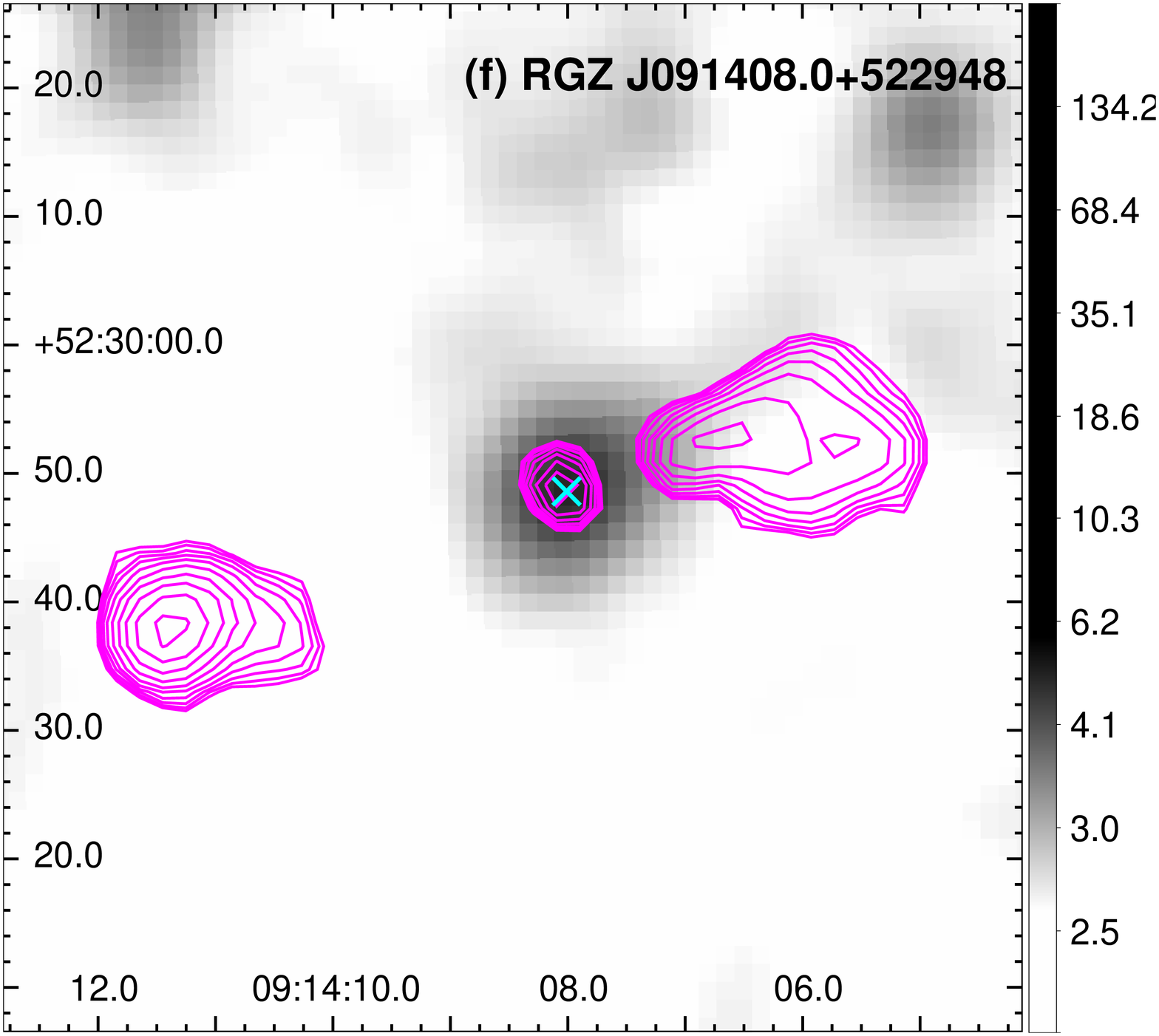}

\includegraphics[width=60mm]{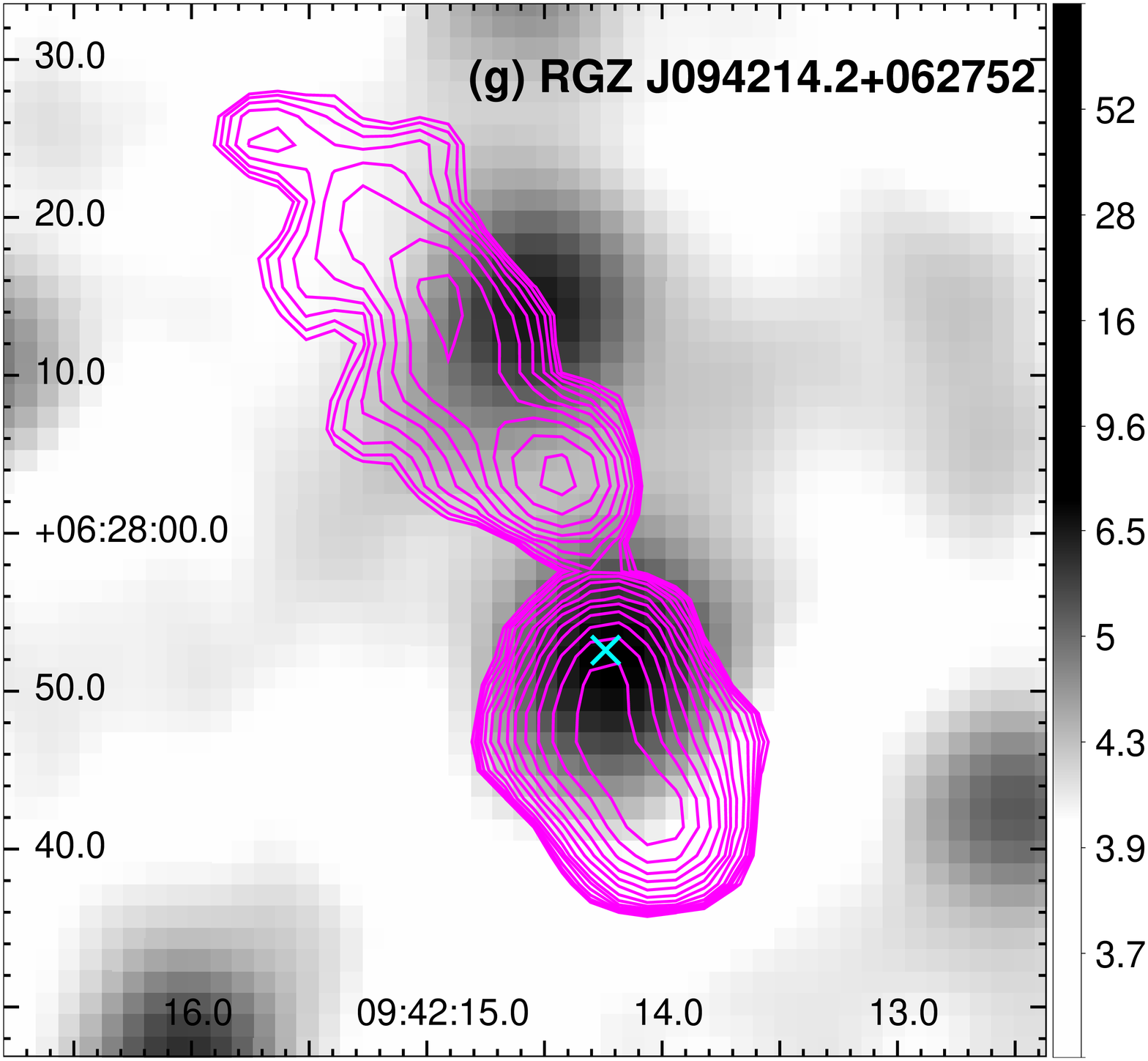}
\includegraphics[width=60mm]{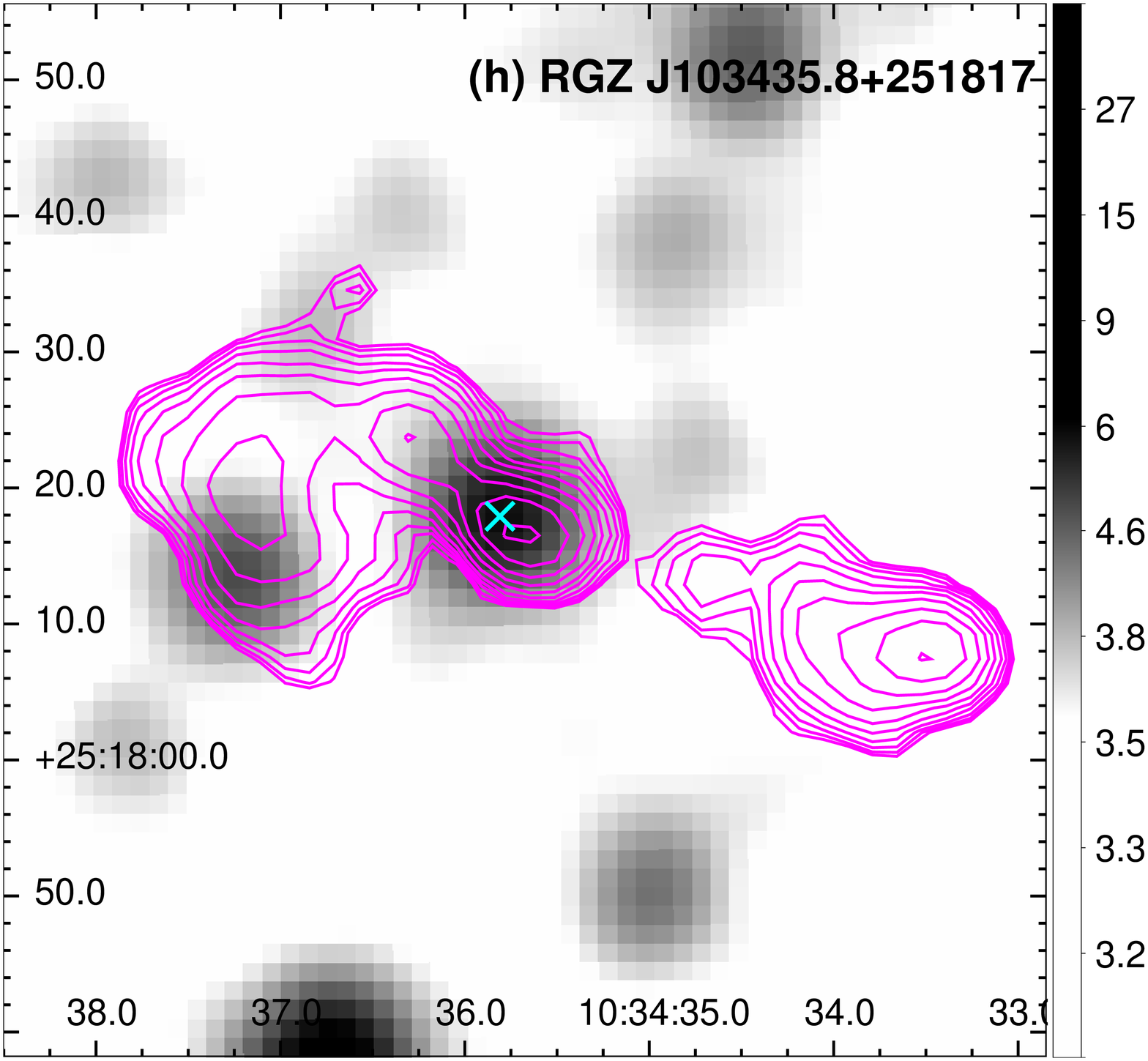}
\includegraphics[width=60mm]{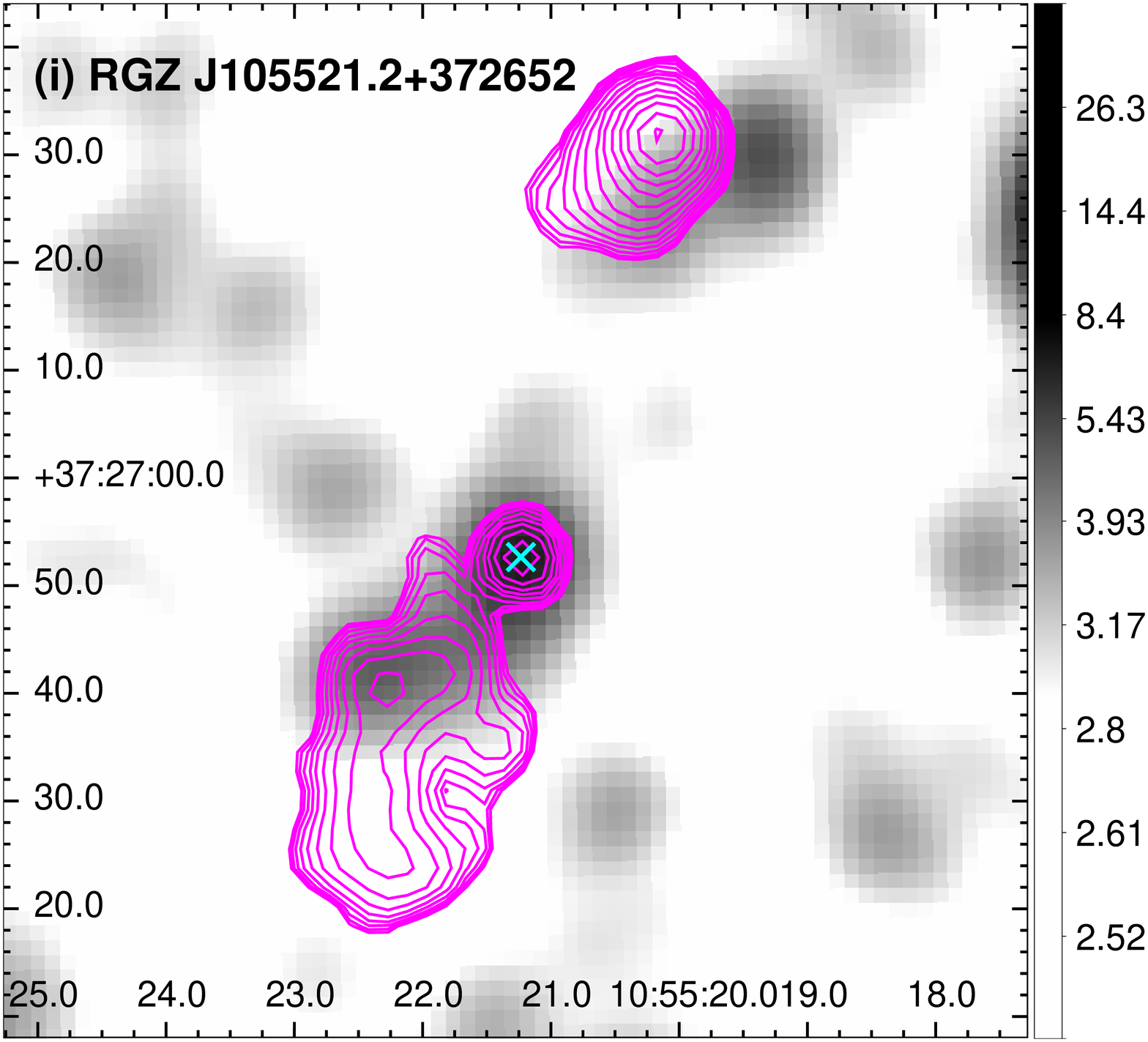}

\includegraphics[width=60mm]{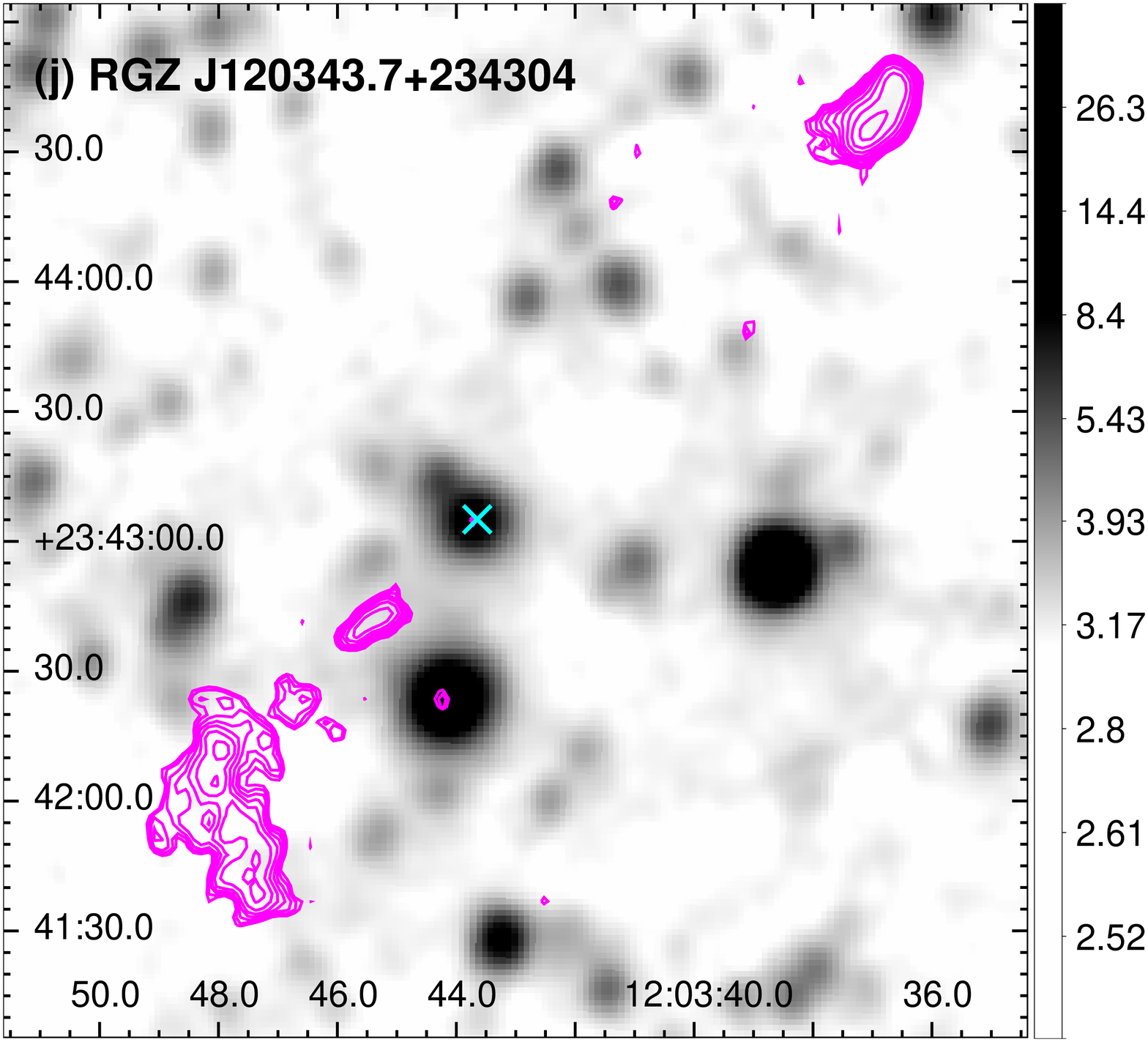}
\includegraphics[width=60mm]{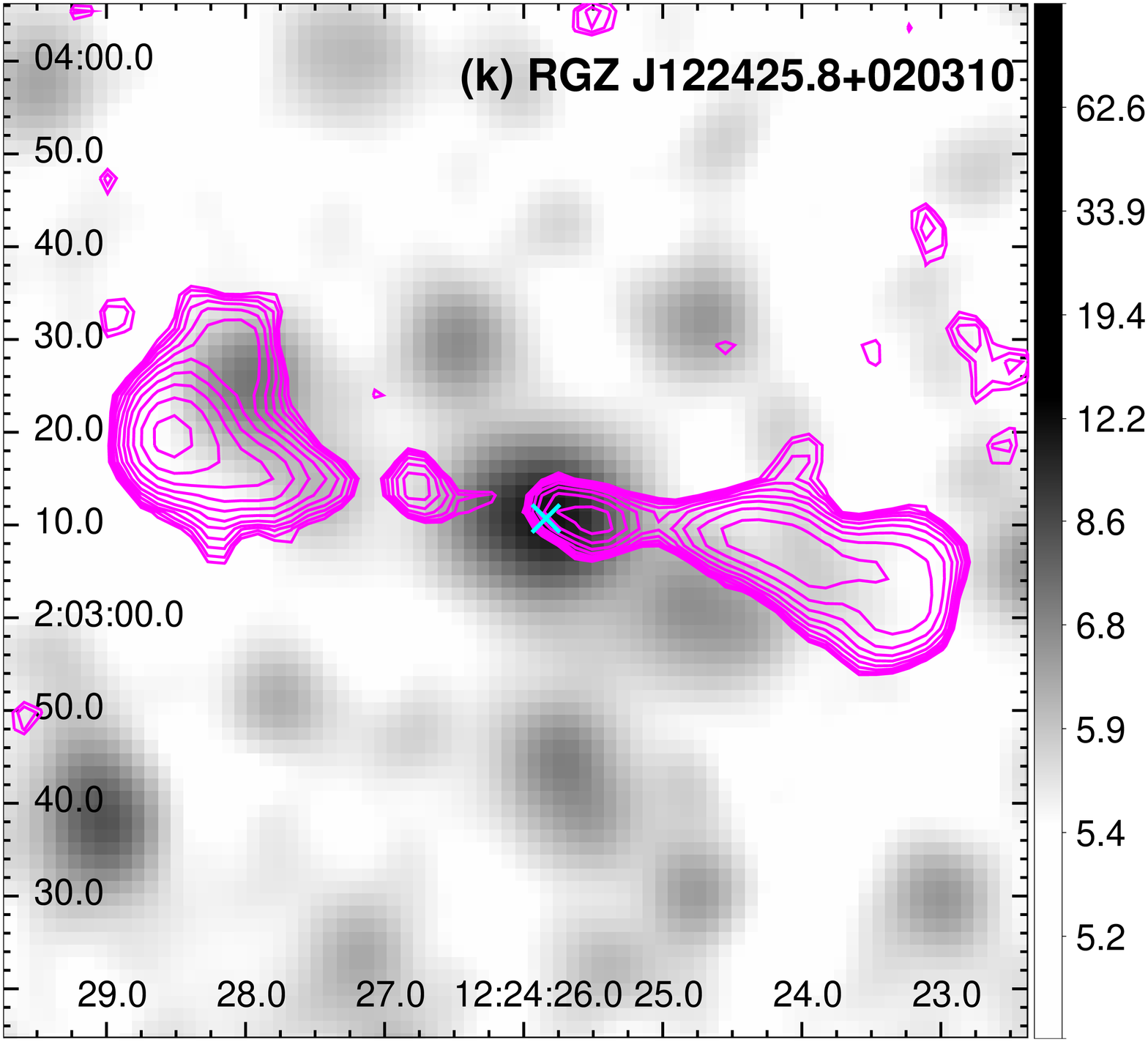}
\includegraphics[width=60mm]{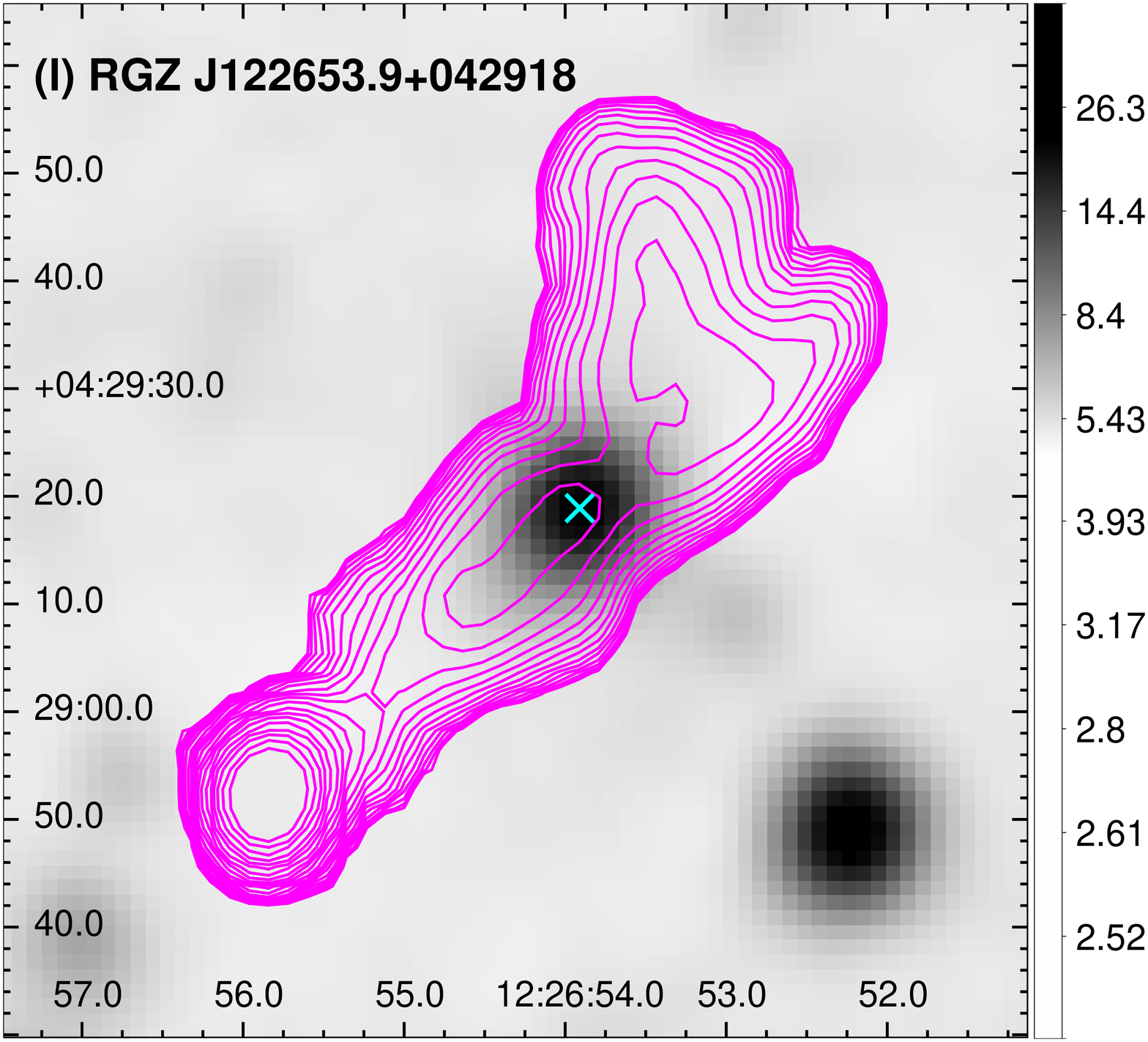}\\

\caption{{\it WISE} $3.4\mu$m images of candidate HyMoRS with overlaid FIRST contours. Hosts of the radio galaxies are marked with crosses. The FIRST contours are drawn at levels $3\sigma+\sigma\times2^{n/2}$ for $n\geq0$, and where $\sigma=0.15$~mJy~beam$^{-1}$. The colourbars are in log scale and are in the raw intensity units of DN/pixel. For RGZ J123300.{2}+06032{5} we additionally include a true color SDSS image of its rare green bean galaxy host (panel m).}
\label{rys:hymors}
\end{figure*}

\setcounter{figure}{1}
\begin{figure*}

\includegraphics[width=120mm]{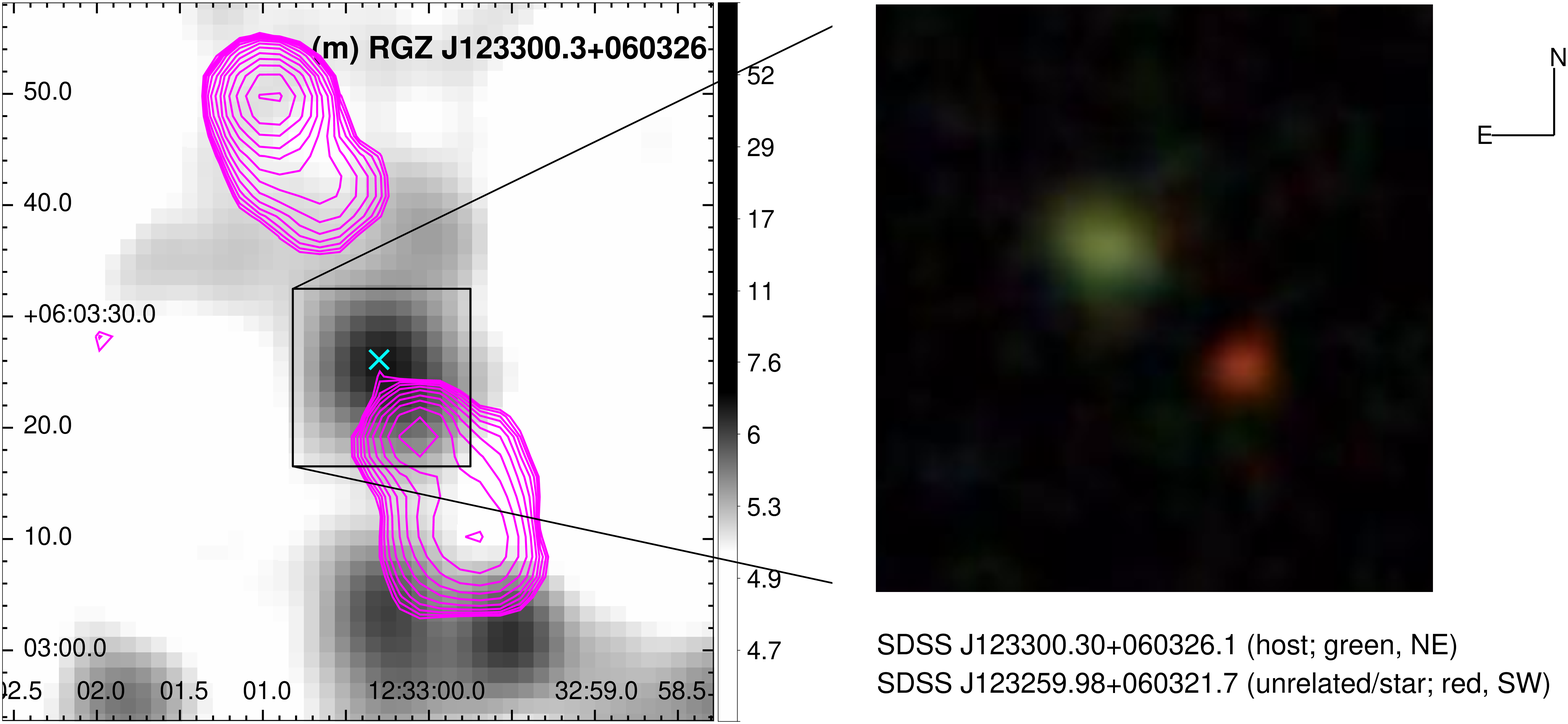}
\includegraphics[width=60mm]{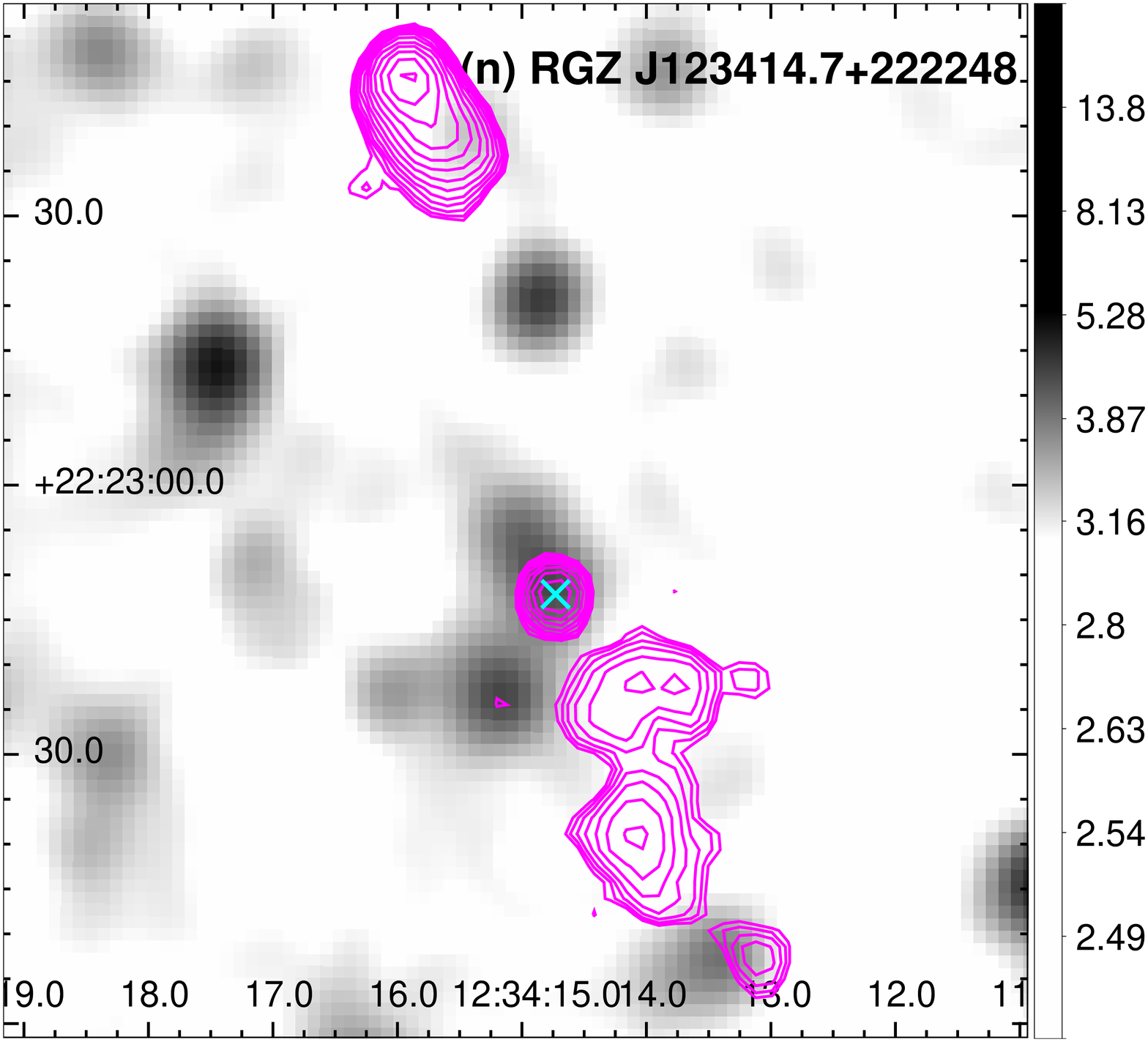}

\includegraphics[width=60mm]{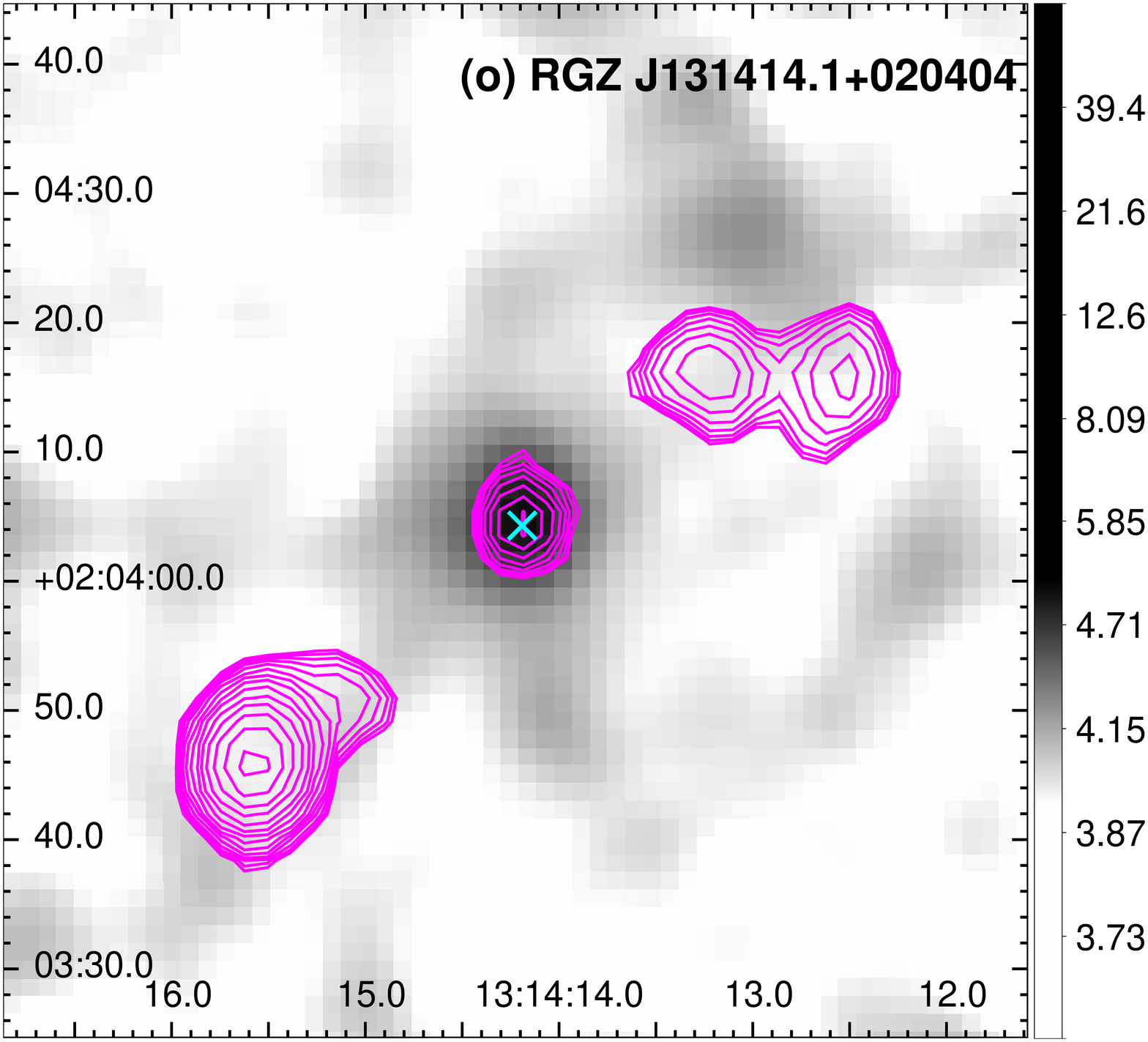}
\includegraphics[width=60mm]{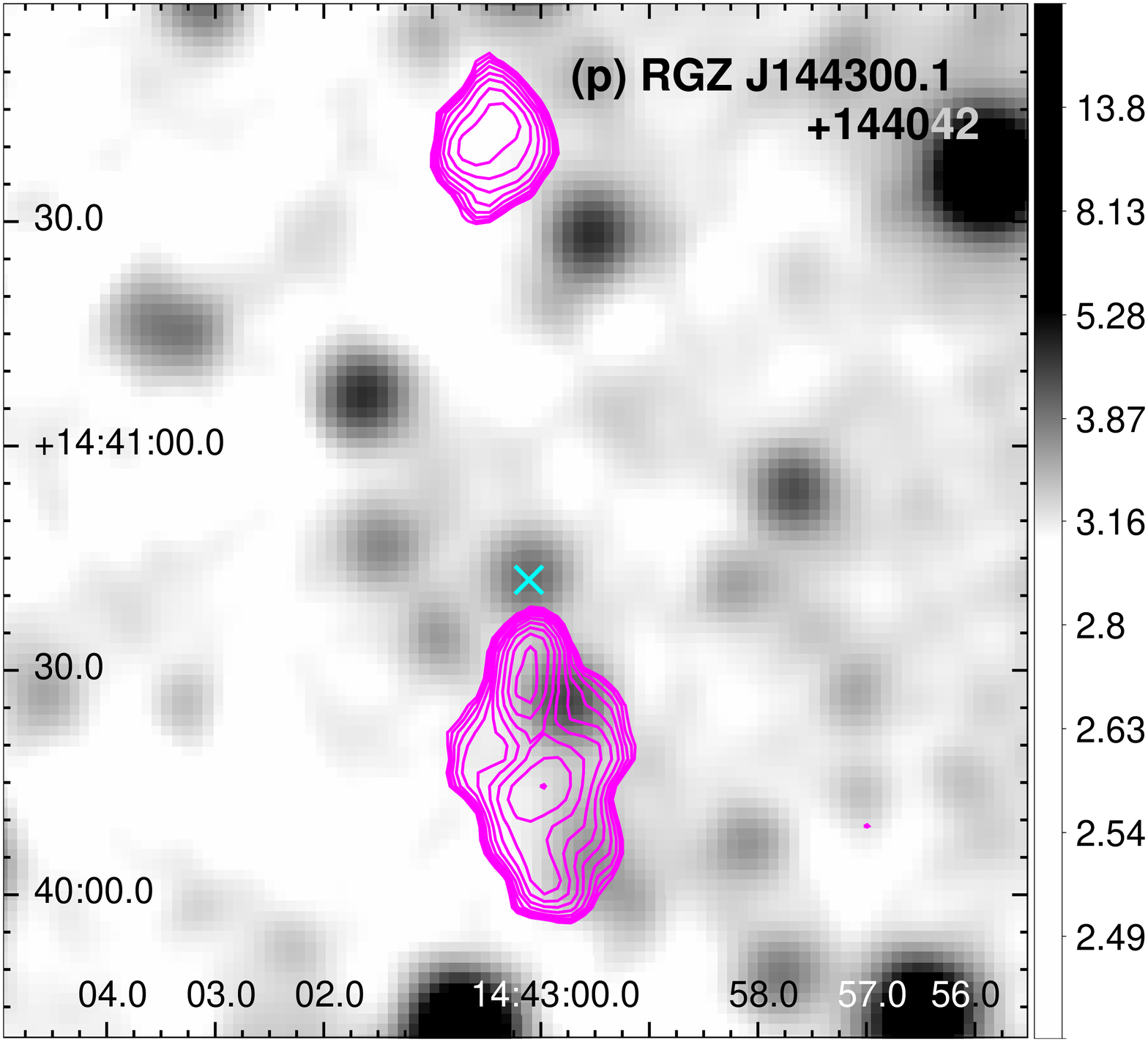}
\includegraphics[width=60mm]{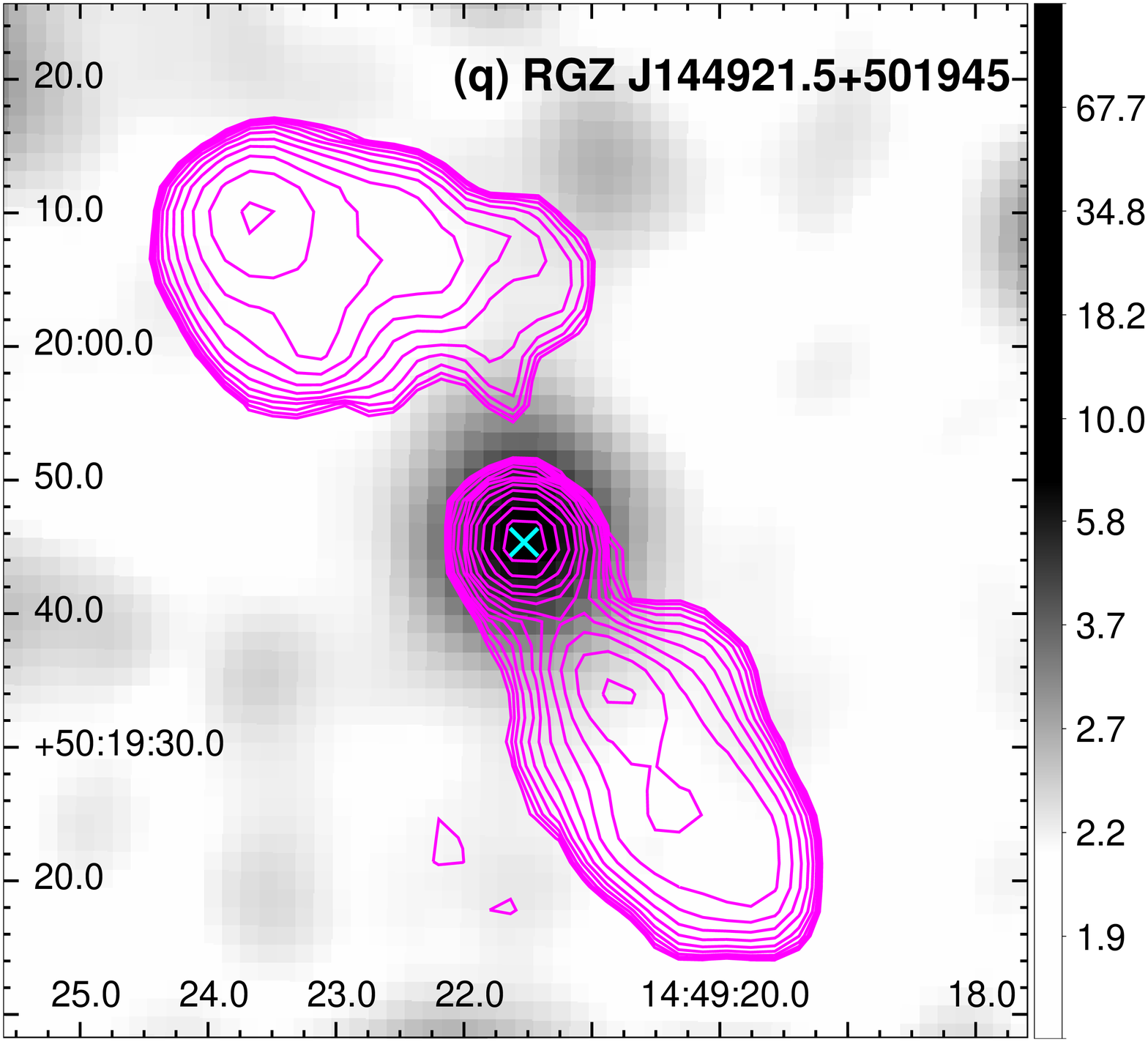}

\includegraphics[width=60mm]{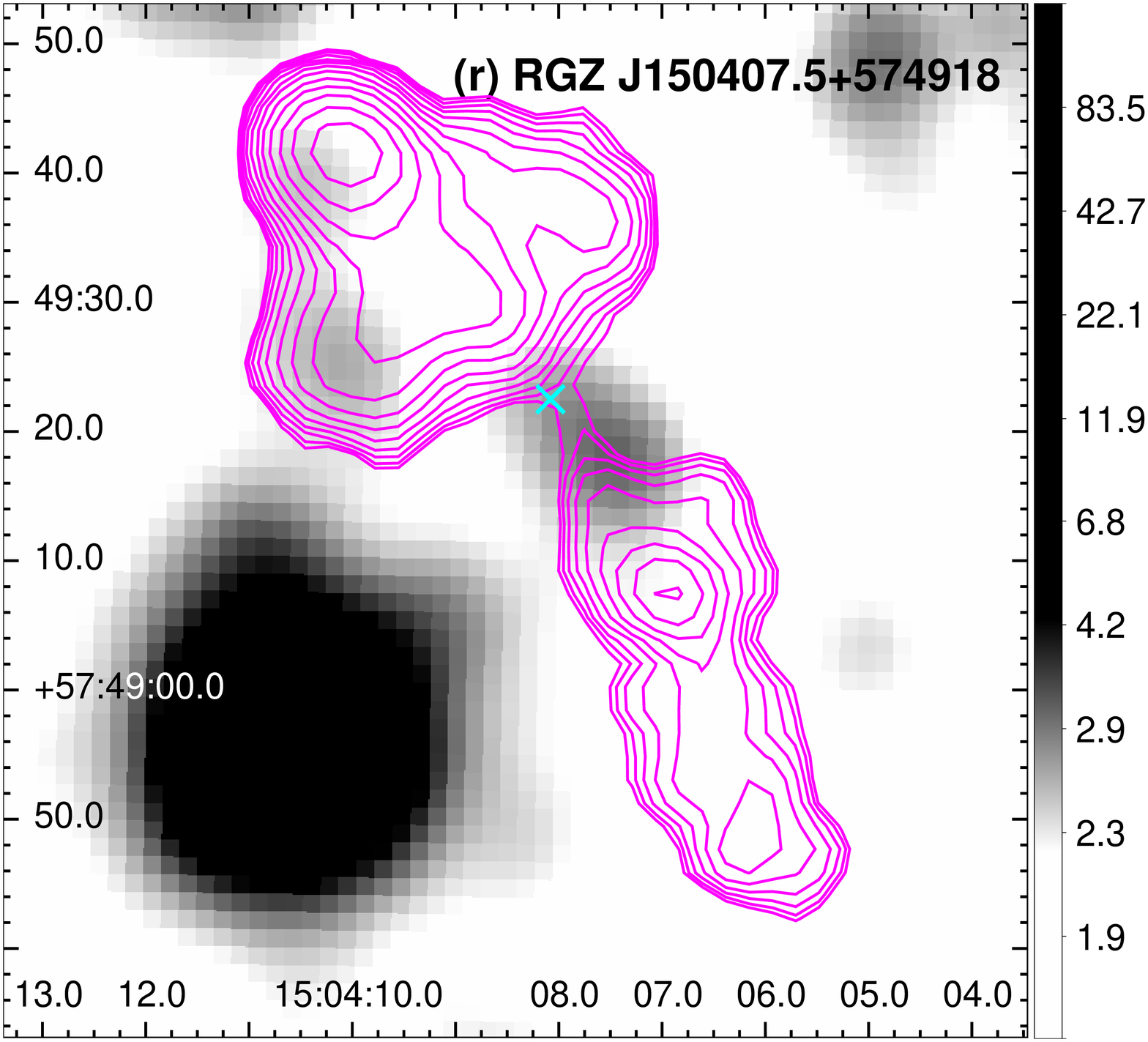}
\includegraphics[width=60mm]{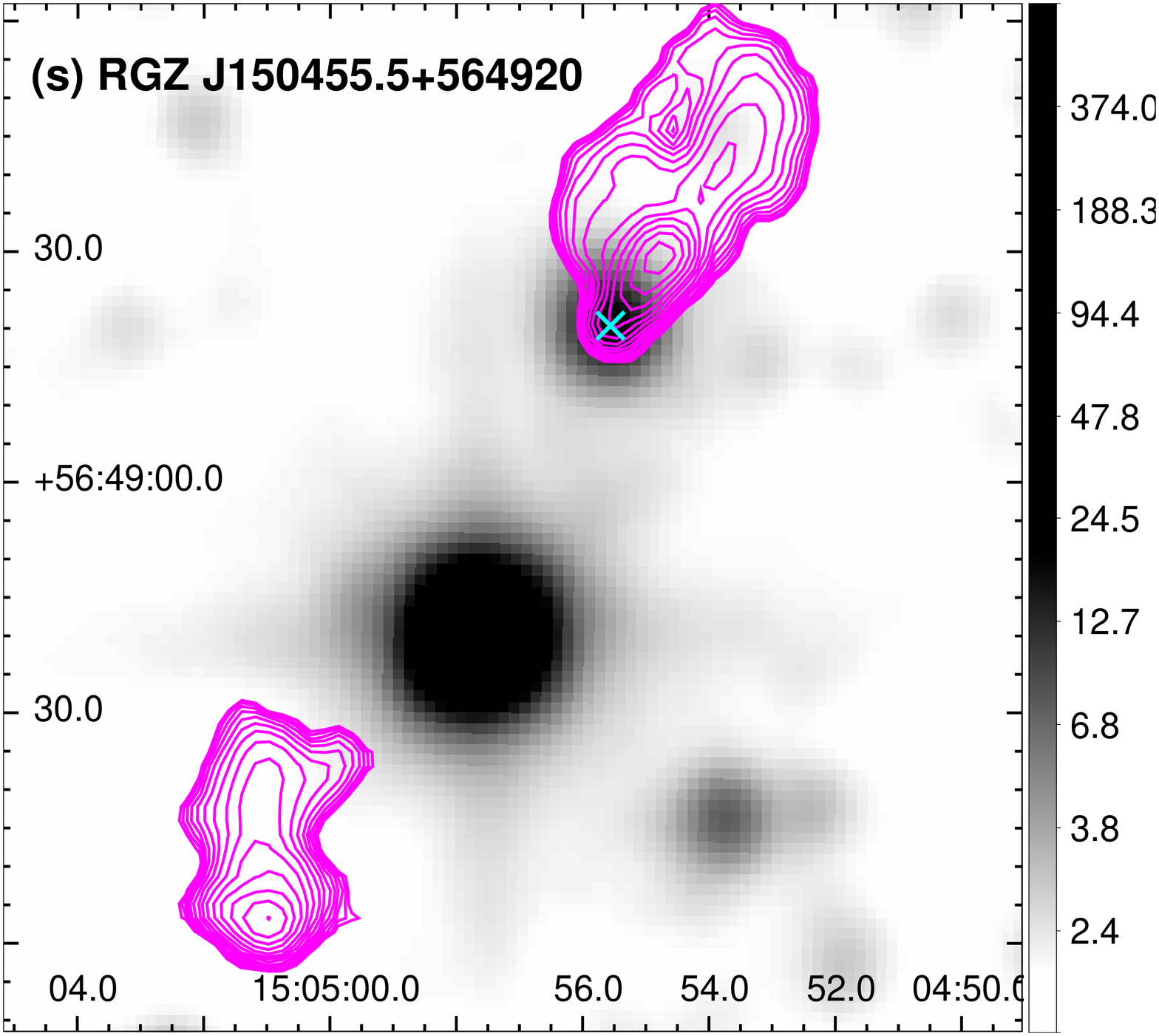}
\includegraphics[width=60mm]{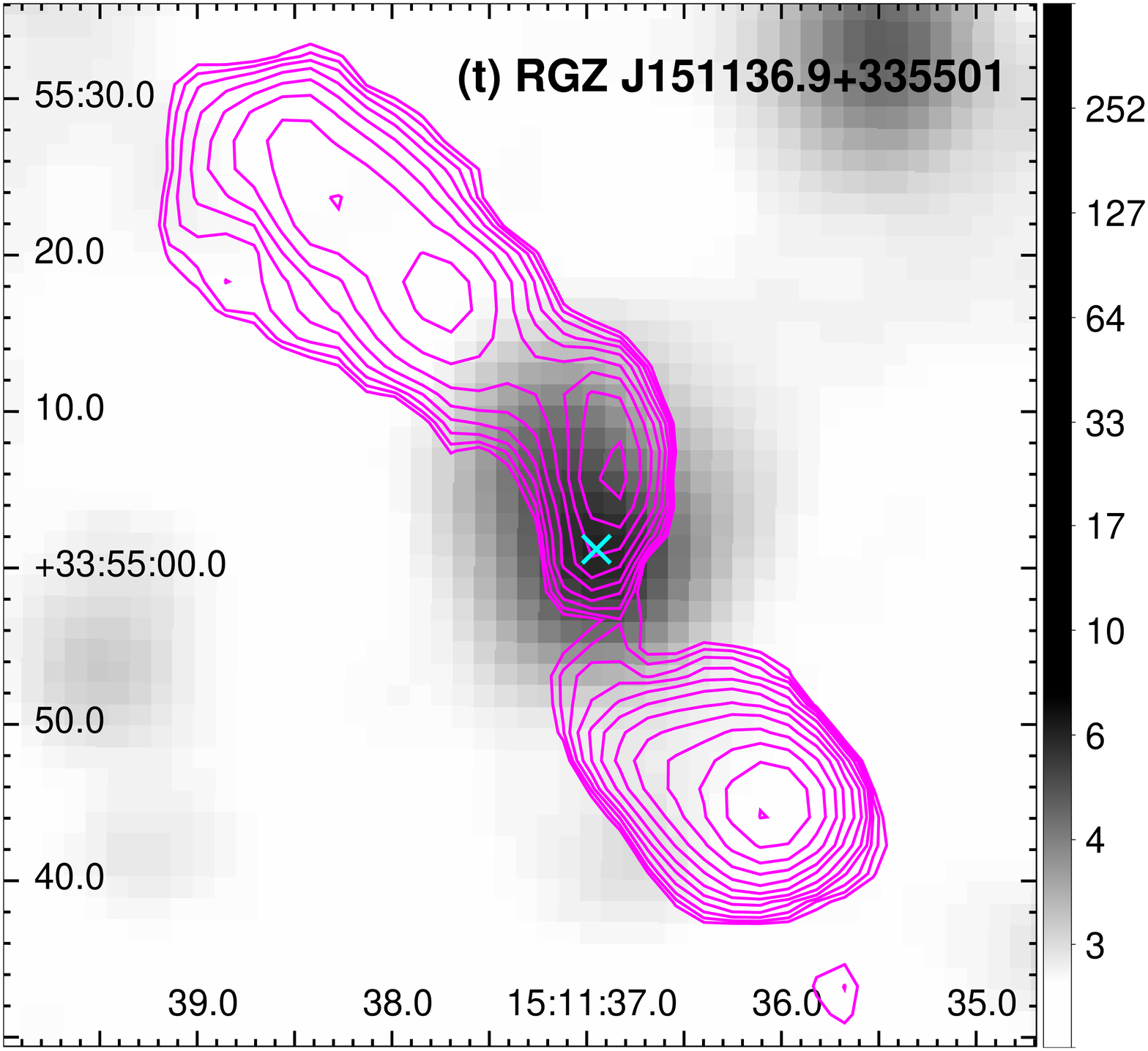}

\includegraphics[width=60mm]{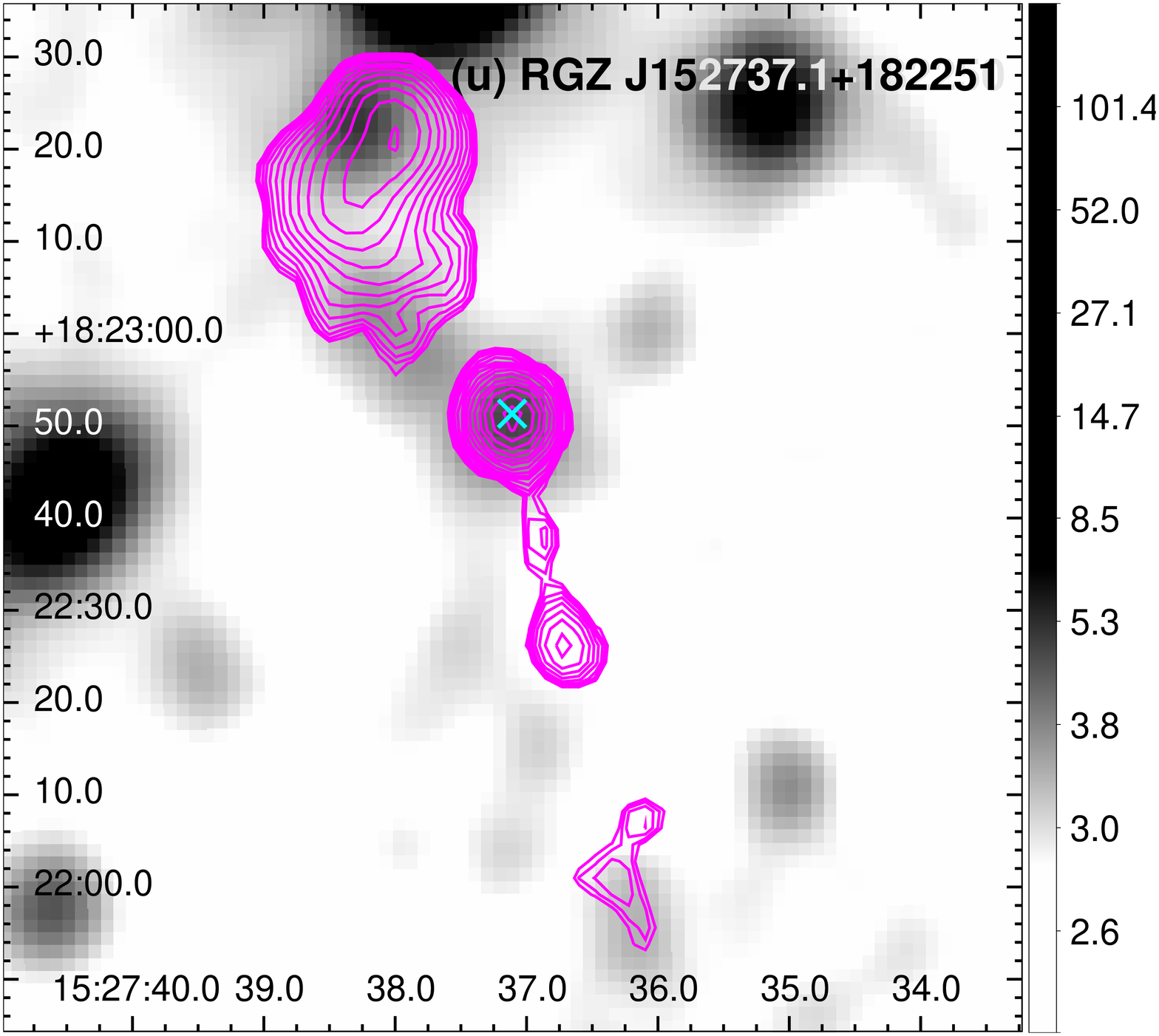}
\includegraphics[width=60mm]{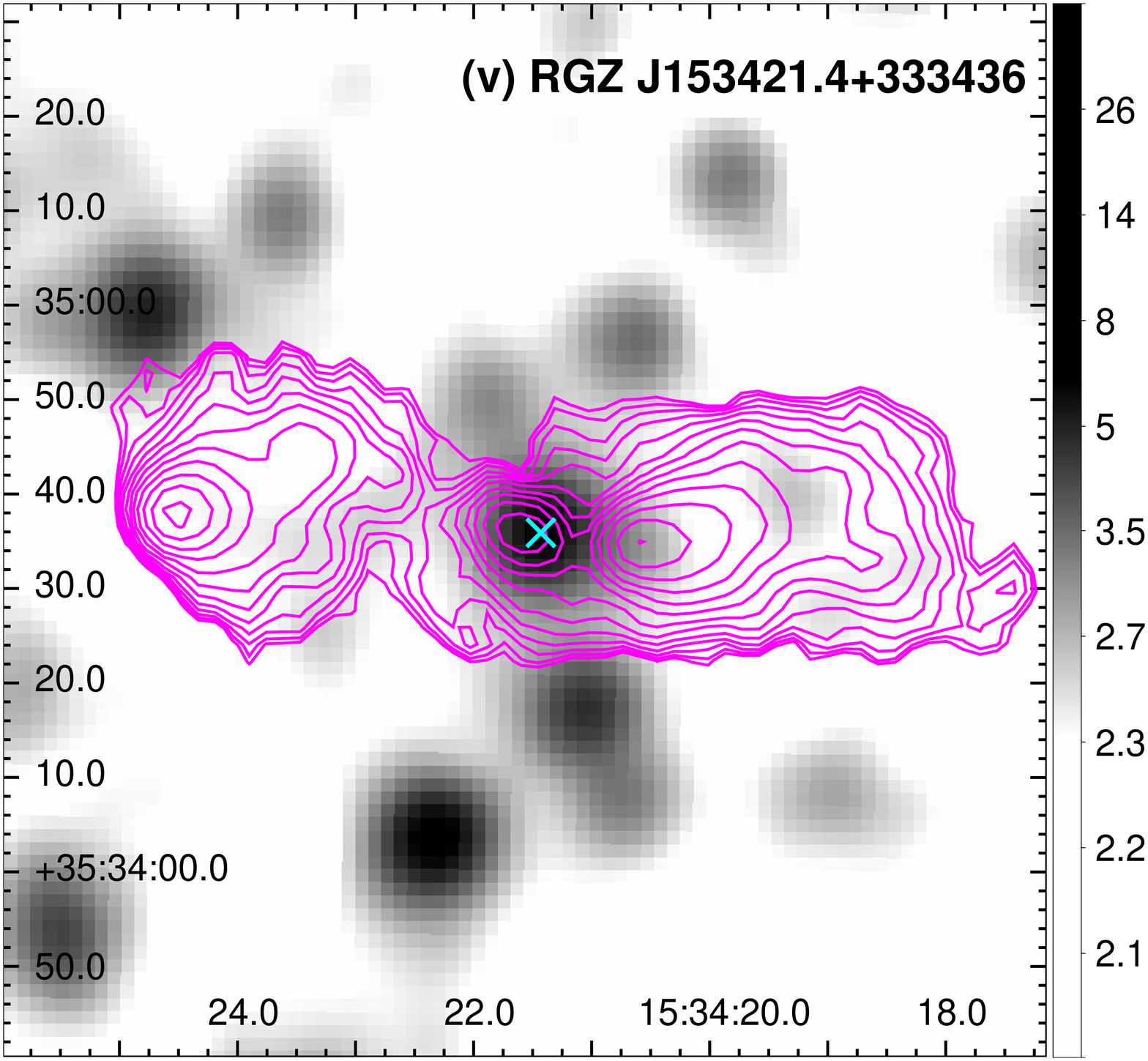}
\includegraphics[width=60mm]{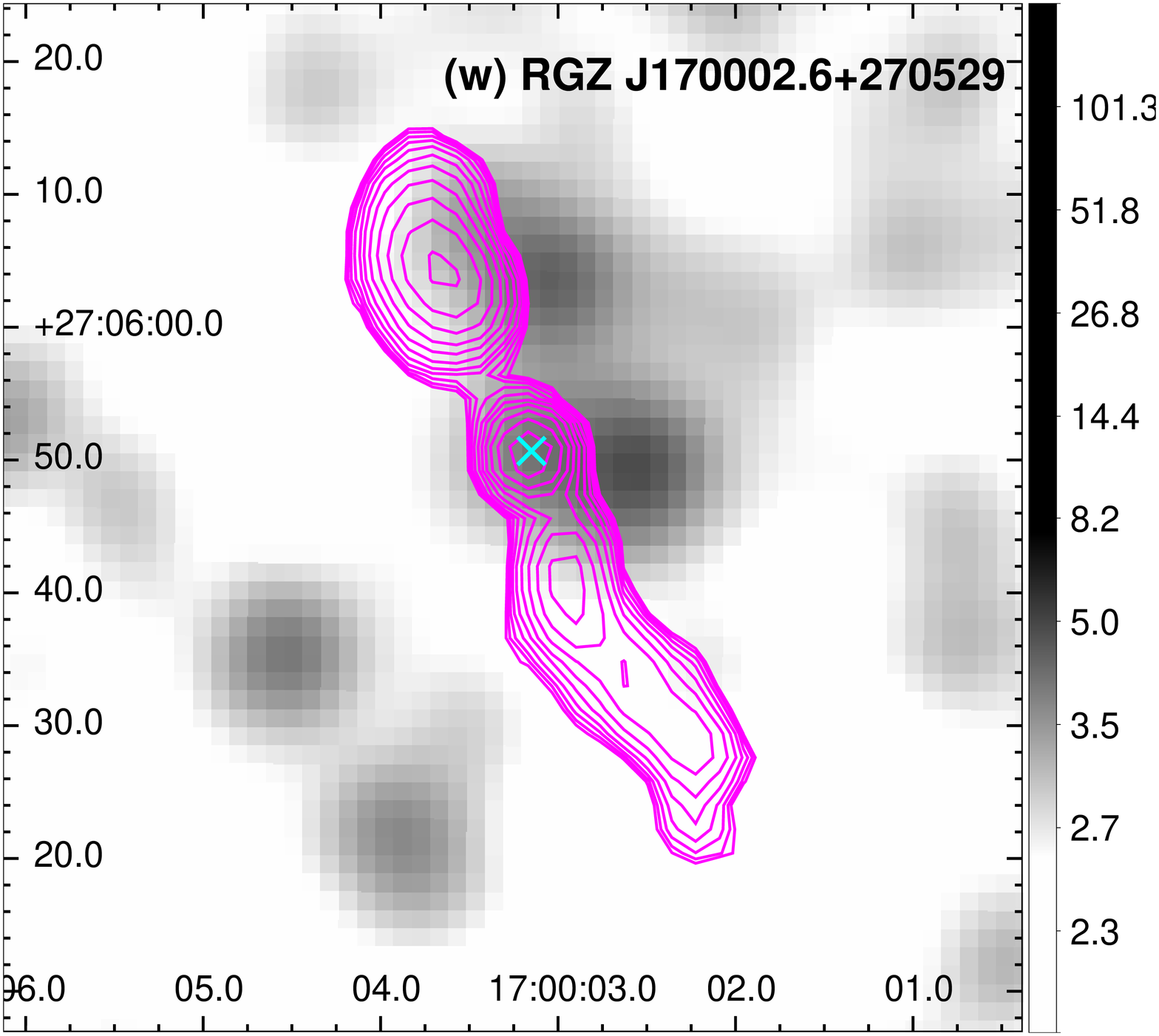}

\caption{\it Continued.}
\end{figure*}


11. {\bf RGZ J105838.6+244535} (Figure~\ref{rys:hymors-large}b): In the FIRST image the NW lobe displays a strong hotspot-like component at its far end. The SE lobe is resolved out in FIRST, but its diffuse emission is detected in the lower resolution NVSS image. There is no compact component in the SE lobe, and its overall structure seems relaxed. A compact radio core is detected in the FIRST image. This is a giant radio galaxy, with a size of 2.3~Mpc \cite[][]{2017arXiv170400516D}. 

12. {\bf RGZ J120343.7+234304} (Figure~\ref{rys:hymors}j): The NW lobe is dominated by a bright, albeit extended at FIRST angular resolution of $5.4''$, hotspot-like component. In the NVSS image no radio emission is detected beyond that detected in FIRST. The SE lobe consists of a pointed tail-like emission that distorts and bends in the outer parts of the lobe. A faint radio core is detected. The host has targeted SDSS spectroscopic observations; located at $z=0.1767$ the galaxy is a low excitation radio galaxy (LERG; see \S\ref{sec:optical}).

13. {\bf RGZ J122425.8+020310} (Figure~\ref{rys:hymors}k): The eastern lobe displays {a} strong hotspot-like component at its far end, with lobe emission pointing back towards the host galaxy. Some of the lobe emission close to the hotspot-like component extends sideways, to the north. The western lobe is devoid of any strong compact component, but it is rather straight becoming more extended sideways at the lobe far end. If a radio core is present, it is merged with the western lobe emission. The host has targeted SDSS BOSS spectroscopic observations; located at $z=0.4516$ the galaxy is a LERG (see \S\ref{sec:optical}). The host of this radio galaxy is the brightest cluster galaxy (BCG) of a galaxy cluster WHL J122425.8+020310 \cite[][this is a new estimate of the cluster redshift, improving on the original photometric estimate]{2009ApJS..183..197W}.

14. {\bf RGZ J122653.9+041918} (4C +04.43; Figure~\ref{rys:hymors}l): The SE lobe is dominated by a {hotspot}-like component. This component has been previously interpreted as a compact radio core, but no optical (SDSS) nor infrared ({\it WISE}) counterparts are found in its vicinity, and no radio variability has been found \cite[][]{1995AZh....72..291G,2011ApJ...737...45O}. The NW lobe displays a bifurcation extending north and north-west. We note this is a re-discovery, {since} the source has been marked previously by \cite{2011ApJS..194...31P} as a possible HyMoRS. If a radio core is present, it is merged with the diffuse lobe emission. The host has targeted SDSS spectroscopic observations; located at $z=0.5174$ it is classified as a QSO. 

15. {\bf RGZ J123300.2+060325} (Figure~\ref{rys:hymors}m): The NE lobe displays a hotspot-like component at its far end, with extended emission pointing back towards the host galaxy. The SW lobe is more relaxed, with no clear compact component. No radio core is detected. 

16. {\bf RGZ J123414.7+222248} (7C 1231+2239; Figure~\ref{rys:hymors}n): The northern lobe is dominated by a strong hotspot-like component. The southern lobe displays diffuse meandering emission. A compact radio core is detected. 

17. {\bf RGZ J131414.1+020404} (Figure~\ref{rys:hymors}o): The SE lobe displays a strong hotspot-like component, with low-level emission pointing towards the radio core. The NW lobe is bent and is of a tail-like structure. A compact radio core is detected.

18. {\bf RGZ J144300.1+144042} (Figure~\ref{rys:hymors}p): The northern lobe displays {a} bright {component, which is} resolved at the {FIRST} angular resolution of $5.4''$. No additional emission is detected in NVSS beyond the FIRST component. The southern lobe displays elongated diffuse emission. No radio core is detected. This source is, to some degree, reminiscent of the HyMoRS reported by \cite{2011BASI...39..547P}.

19. {\bf RGZ J144921.5+501945} (Figure~\ref{rys:hymors}q): The NE lobe displays a strong hotspot-like component at its far end, with lobe diffuse emission extending both towards the radio core and sideways{, perhaps representing a backflow}. The SW lobe has no hotspot-like components, and its brightest peak is located close to the radio core. A compact radio core is detected. This radio galaxy has been previo{us}ly classified as a possible BL Lac object \cite[][]{2014ApJS..215...14D}.

20. {\bf RGZ J150407.5+574918} (Figure~\ref{rys:hymors}r): The NE lobe displays a hotspot-like component at its far end, and more diffuse lobe emission pointing towards the position of the host galaxy, but also extending somewhat sideways. The SW lobe is of a plume-like, relaxed and meandering structure. No clear radio core is detected, but the host is most likely SDSSJ150408.08+574922.5. The mid-IR counterpart is confused with a nearby galaxy.

21. {\bf RGZ J150455.5+564920} (Figure~\ref{rys:hymors}s): The SE lobe displays a strong hotspot-like component, with extended emission bent and pointing towards the direction of the host galaxy. The NW lobe is of more extended, diffuse and relaxed structure, with brightness peak occuring close to the position of the host galaxy. If {a} radio core is present, it is merged with the NW lobe emission. The host has targeted SDSS spectroscopic observations; located at $z=0.3587$ it is classified as a broadline QSO. 

22. {\bf RGZ J151136.9+335501} (7C 1509+3406; Figure~\ref{rys:hymors}t): The SW lobe displays a hotspot-like component with diffuse lobe emission pointing eastwards. The NE lobe is of a plume-like structure. A radio core is not easily distinguishable, and if present it is merged with the diffuse emission connecting the lobes of the radio galaxy. The host has targeted SDSS spectroscopic observations; located at $z=0.6234$ it is classified as a QSO. 

23. {\bf RGZ J152737.1+182250} (Figure~\ref{rys:hymors}u): The {NE} lobe displays a hotspot-like component at its far end, with extended emission bending and extending south, but slightly pointing towards the radio core. The SW lobe features a brightness peak in the proximity of the srong radio core, with some detached emission further away. A strong, compact radio core is detected. 

24. {\bf RGZ J153421.4+333436} (7C 1532+3344; Figure~\ref{rys:hymors}v): The eastern lobe features a strong hotspot-like component at its far end, with diffuse lobe emission extending between the hotspot and the radio core. The western lobe features a brightness peak close to the radio core, and displays diffuse emission extending far beyond the brightness peak, away f{ro}m the radio core. A radio core is detected, but is embedded in the diffuse lobe emission. 

25. {\bf RGZ J170002.6+270549} (Figure~\ref{rys:hymors}w): The NE lobe features a single, but somewhat extended strong component. The SW lobe is devoid of compact components and {consists} of a tail-like structure slightly meandering away from the radio core. A radio core is detected.


\subsection{Radio properties}
\label{sec:radio}

We measure the total radio luminosity density at 1.4~GHz, the radio spectral index between 200~MHz and 4.85~GHz, and the total projected linear extent of each of our candidate HyMoRS. Results are presented in Tables~\ref{tab:optical} and \ref{tab:wise-radio}. The radio luminosity densities are k-corrected using the measured radio spectral index (unless unavailable, in which case the canonical $\alpha=0.75$ is used). The total projected linear extent is measured between the outermost $3\sigma$ contours in the FIRST images \cite[][]{1995MNRAS.277..331S}. The results are plotted in Figure~\ref{rys:radio-opt}, and further discussed in \S\ref{sec:disc-radio}. 

{Although our candidates remain selected solely on the  basis of visual inspection, t}o attempt a quantitative assessment of radio morphology of the candidates in this paper, we {calculate the} $f_{\rm FR}$ index \cite[][]{2012MNRAS.427.3196K}  that stems directly from the original \cite{1974MNRAS.167P..31F} FR definition
\begin{equation}
f_{\rm FR} = \frac{2 x_{\rm bright}}{x_{\rm total}} + 0.5,
\label{eqn:FRindex}
\end{equation}
where $x_{\rm bright}$ is {the} distance of the brightest pixel in the lobe from the position of the host, and $x_{\rm total}$ is the distance between the position of the optical host and farthest extent of the lobe. Following \cite{2012MNRAS.427.3196K}, if $0.5 \leqslant f_{\rm FR} \leqslant 1.5$ the lobe is of an FRI morphology, and if $1.5 < f_{\rm FR} \leqslant 2.5$ the lobe is {considered that} of an FRII morphology. There are a few caveats associated with the use of this definition, however. For example, the index value will be overestimated if the extended emission in the outer parts of the lobe is resolved out due to shortcomings of the observations. {T}he FIRST survey is particular{l}y susceptible to this effect {because of the lack of short baselines}. Also, this method works best when a lobe hosts a clear brightness peak; however, the index may be misleading when the surface brightness {emission} of the lobe displays little variation with distance from the core. Therefore, the index should be considered an indication rather than a strict classification method. We measure the $f_{\rm FR}$ index for each lobe of each HyMoRS candidate. We use the FIRST survey for the $x_{\rm bright}$ and $x_{\rm total}$ measurements (unless otherwise stated in the results table) and the results are presented in Table~\ref{tab:radio-ratio}.

The FR~II sides of all candidate HyMoRS in this paper are quantified as such with Eqn.~\ref{eqn:FRindex} ($f_{\rm FR}>1.5$). In the case of the FR~I sides four candidates show index values of $1.68\leq f_{\rm FR}\leq2.02$, which in principle classifies them as of {an} FR~II morphology. In the case of RGZ~J072406.{7}+38034{8} (Figure~\ref{rys:hymors}a; $f_{\rm FR}=2.02\pm0.07$ for the FR~I side) and RGZ~J084738.{0}+183156 (Figure~\ref{rys:hymors}d; $f_{\rm FR}=1.79\pm0.17$ for the FR~I side) there is a possibility that faint low surface brightness {emission} extending further away is resolved out in the FIRST images, causing $x_{\rm total}$ to be underestimated. RGZ~J122425.8+02031{0} (Figure~\ref{rys:hymors}k; $f_{\rm FR}=1.78\pm0.10$ for the FR~I side) displays a plateau-like emission (as opposed to peaked) across more than 50\% of the lobe extent. We also note that given the uncertainties the FR indices of RGZ~J084738.{0}+183156, RGZ~J122425.8+02031{0} and RGZ~J131414.{1}+02040{4} (Figures~\ref{rys:hymors}{d, 2k, 2}o) are all borderline cases.

{Although the $f_{\rm FR}$ formula is based on the definition of FRI and FRII morphology, we additionally verified the range of the index values for typical FRIs and FRIIs. We selected extended radio galaxies from the 3CRR catalog \cite[][]{1983MNRAS.204..151L} that were located at redshifts $z<1.0$ and within the FIRST survey coverage. The 3CRR radio galaxies are powerful radio sources, but weaker sources have not been studied as extensively and radio morphology classification of 3CRR radio galaxies is very secure. For consistency, we use the FIRST survey images to measure the $f_{\rm FR}$ index of the 3CRR sources. We find that the average $f_{\rm FR}$ index for lobes of FRI radio sources is $1.13\pm0.29$, with a median of $1.17$ (6 radio sources, 12 measurements). There is only one lobe of one FRI radio galaxy that is an outlier with $f_{\rm FR}=1.57\pm0.13$. For FRII radio sources the average $f_{\rm FR}$ index is $2.13\pm0.28$, with a median of $2.24$, and all lobes of FRII sources have $f_{\rm FR}>1.49$ (11 radio sources, 22 measurements).}

\subsection{Optical properties}
\label{sec:optical}

{The absolute magnitudes in the optical $R$ band are rest-frame \cite[k-correction performed using the online calculator\footnote{\tt http://kcor.sai.msu.ru/} of ][]{2010MNRAS.405.1409C,2012MNRAS.419.1727C}.  S}pectroscopic observations are publicly available for {nine} candidates, as detailed in Table~\ref{tab:optical} and \S\ref{sec:hymors-notes}. {Six} hosts are quasars as classified in the SDSS database {and \cite{1994AJ....107.1245S}}. We measure emission lines of the remaining three hosts to classify them as either high- or low-excitation galaxies using  the so-called excitation index $EI$ \cite[][]{2010A&A...509A...6B}, where
\begin{multline}
EI = \text{log}_{10} \left( \frac{\text{[O\sc{iii}]}}{\text{H}\beta}\right) -  \frac{1}{3} \times \\ \left[
     \text{log}_{10} \left( \frac{\text{[N\sc{ii}]}}{\text{H}\alpha}\right) 
   + \text{log}_{10} \left( \frac{\text{[S\sc{ii}]}}{\text{H}\alpha}\right)
   + \text{log}_{10} \left( \frac{\text{[O\sc{i}]}}{\text{H}\alpha}\right)       \right].
\end{multline}

\vspace{2mm}
\noindent
If $EI>0.95$ the galaxy is classified as high excitation, otherwise as low excitation \cite[][]{2010A&A...509A...6B}.

We find the excitation indices $EI=1.44$ (HERG) for RGZ J103435.8+25181{7},  and $EI=0.50$ (LERG) for RGZ J122425.8+02031{0}. In the case of RGZ J1203 43.7+234305 the [N{\sc ii}], [S{\sc ii}] and H$\alpha$ lines are not available, hence we use a simplified classification method from \cite{2012MNRAS.421.1569B}, which is based purely on the equivalent width of the [O{\sc iii}] line (EW). Specifically, \cite{2012MNRAS.421.1569B} found that for $EW(\text{[O\sc{iii}]})<5$\AA ~the galaxy is most likely low excitation. Likewise, \cite{1998MNRAS.298.1035T} reported $EW(\text{[O\sc{iii}]})>10$\AA ~for HERGs. We find $EW(\text{[O\sc{iii}]})=2.19\pm0.58$\AA ~for RGZ J120343.7+23430{4}, and hence formally classifiy it as LERG. We also note that the spectrum of this galaxy in general lacks any strong emission lines, what strenghtens its classification as a LERG. All line fitting parameters are taken from the SDSS DR13 database.

We also investigate in more detail the optical host of RGZ~J123300.{2}+06032{5}. The host is most likely a rare green bean galaxy \cite[GBG;][]{2013ApJ...763...60S}. The GBGs are extended objects (Petrosian radius $r_{\rm Petro}>2''$) and have $g-r>1.0$. The host of RGZ J123300.{2}+06032{5} has $r_{\rm Petro} = 2.61''\pm0.17''$, and with colours $g-r=1.02\pm0.04$, $r-i=0.06\pm0.03$, $u-r=2.36\pm0.20$ and $r-z=0.57\pm0.05$ it is located in the region occupied by galaxies in Figure~2 of \cite{2009MNRAS.399.1191C}, outside the selection requirement for green {bean} galaxies. The host colours meet 11 of 12 selection criteria for selection of GBGs proposed by \cite{2013ApJ...763...60S}. We discuss this further in \S\ref{sec:disc-quasars}.

\newpage
\subsection{Citizen scientists' success rates}
\label{sec:hymors-cat}

Fifteen candidates are included in the RGZ DR1 catalog (Wong et al. {\it in prep.}; excluding the giant radio galaxies, which we deemed too difficult {given the RGZ design since} the large angular size of these two radio galaxies that extend well beyond the $3'\times3'$ cut-out). Our candidates are all multi-component radio sources at the FIRST survey angular resolution, and for this reason they can appear multiple times in the RGZ catalog (one entry for each radio component). We assess each entry separately using two criteria: one, if the mid-IR host of the candidate HyMoRS was correctly identified, and two, if all radio components have been assigned to the overall radio structure of the source. Results are presented in Table~\ref{tab:rgzcat}.

For 12 candidates all radio components have been correctly assigned by the citizen scientists (80\% success rate). For 12 candidates the mid-IR hosts have been identified correctly by the citizen scientists (80\% success rate). For 11 candidates the citizen scientists corrently assigned all radio components and identified the mid-IR host at the same time (73\% success rate). Candidates with the multiple entries in the RGZ  DR1 catalog, that have inconsistent classifications between the entries, are considered ambigous classifications (13\%).

Complete radio component selection proved to be difficult for the citizen scientists in the case of multi-component radio sources with angular sizes of $\gtrsim 115''$. For radio galaxies of angular sizes $\leqslant 110''$ the success rates are 100\% for the radio component association, and 90\% for the mid-IR host selection (10\% ambiguous). This result indicates that the main obstacle for the citizen scientists might have been the limited image size of the cut-outs they were presented {with}. For example, in the case of RGZ~120343.7+23{43}05 ($4.25'$ angular size, 785~kpc linear size) the citizen scientists were presented with only the SE lobe, and as such they could not select the correct host nor identify all radio components of the whole radio galaxy. They did, however, correctly identify all lobe components of the SE lobe and correctly assigned no mid-IR host to it. Unfortunately, for any radio galaxy with angular size exceeding the RGZ cut-out size this will almost always be the case, and thus the users of the RGZ catalog should be aware of this caveat to an otherwise valuable resource.


\section{Discussion}
\label{sec:disc}
\label{sec:origin-props}

\subsection{The origin and formation of HyMoRS}

The origin of the radio structure of HyMoRS is yet to be established. As favoured by current consensus \cite[e.g.][]{2009AN....330..184B,2015Turner}, radio morphology of radio galaxies is due to a combination of the radio source environment (nurture, {\S\ref{sec:nurture}}) and jet power (nature, {\S\ref{sec:nature}}).

\subsubsection{Nurture}
\label{sec:nurture}

As {discussed} in {\S\ref{sec:intro}}, there has been a wealth of study {of} the FR dichotomy of radio galaxies. HyMoRS have been invoked as evidence for the significant environmental impact on the formation of radio morphology of radio galaxies \cite[e.g.][]{2002NewAR..46..357G,2008A&A...491..321M}. 

It has been suggested previously that radio galaxies may initially start as FRII radio sources \cite[e.g.][]{2007MNRAS.381.1548K}, in which case {an} FRI morphology would form through the disruption and deceleration of the jets \cite[][among others]{1994ASPC...54..227L,2008A&A...491..321M,2011MNRAS.418.1138W,2012A&A...545A..65P,2015Turner}. Processes such as entrainment \cite[]{2011MNRAS.418.1138W} and  helical instability \cite[]{2012A&A...545A..65P} caused by interaction of lobes/jets with the surrounding media, and {by} density jumps in a non-uniform external medium \cite[]{2008A&A...491..321M} have been considered. Apart from \cite{2008A&A...491..321M}, who focused specifically on theoretical modelling of HyMoRS, most of the deceleration models have been developed for a purpose of general discussion of the transition of FRII radio sources into FRIs. {However, i}n the case of HyMoRS one needs to observe different radio morphologies on each side of a radio galaxy core. Assuming that the radio power of each of the twin jets is exactly the same, the deceleration would have to occur only on one side of the nucleus of the radio galaxy, indicating a highly asymmetric environment. Such asymmetric environments around radio galaxies have indeed been observed. For more discussion on the density asymmetry in the environments of radio sources see \S\ref{sec:disc-quasars}.

\subsubsection{Nature}
\label{sec:nature}

The observed morphology of a radio source in a two dimensional image may suffer from {\it persistent} projection effects, causing the observed radio structures to represent projected and not intrinsic structures. Such effects include Doppler boosting \cite[e.g.][]{1979ApJ...232...34B,1988gera.book..563K,1982MNRAS.200.1067O,1998MNRAS.296..445H,2002JApA...23..235U}, which may be particularly severe if the radio source is observed at a narrow angle to the line of sight. 

The asymmetry in lobe leng{th}s is sometimes used in the quantification of the projection angle, although we note that this will work only under a strong assumption that the jets are not intrinsically bent nor shortened due to differences in the environments on the twin jet paths. We find that 16\% of candidates show no significant asymmetry in their lobe lengths, for 32\% the longer arm (typically considered far side) is of FRI morphology, and for 52\% the longer arm is of an FRII type. Furthermore, projection effects of intrinsically curved jets (e.g. in wide-angle tail radio galaxies)  may also make sources appear asymmetric even without the presence of Doppler boosting. In this case, the hotspot of the near side may be projected to appear as knots or flare points, hence being classified as of FRI morphology, while the far lobe may appear as an FRII. The {intrinsic} asymmetry of lobes may be difficult to verify in the total intensity radio images of the radio sources. As shown by \cite{2017MNRAS.tmp..211D}, radio spectral maps and polarisation imaging along the radio morphological classification are crucial in the confirmation of HyMoRS candidates and their intrinsic asymmetries.

HyMoRS can also be an intrinsically {\it transient} phenomenon, where it has been postulated that effects involving a combination of central engine modulation and differential light travel time between the approaching and receding parts of the radio source can shape the observed radio morphology \cite[][]{1996AA...316L..13G}. For example, \cite{2012AA...544L...2M} developed a simple model that can explain the observed morphological asymmetry in terms of the re-started activity of some radio galaxies [scenario (a)]. Specifically, assuming that the radio source is at least at a moderate inclination to the observer's line of sight ($\lesssim45^{\circ}$), and the jet production is restarting within $1-80\times10^4$~yrs, the differential light travel time effects may cause an apparent FRII morphology on one side (old), and FRI on the other (new), temporarily forming a HyMoRS from the observer's point of view. At least three {known} HyMoRS can be explained with this model based purely on their radio morphology \cite[e.g. J1211+743, 3C249.1, 3C 334;][]{2012AA...544L...2M,2012A&A...545A.132M}. Although the model was developed to cast doubt on hybrid radio morphology as an intrinsic property of some of the sources, perhaps at least a fraction of the HyMoRS population can be interpreted as a class of AGN transients \cite[but {see} ][]{2013A&A...557A..75C}.

Here we also suggest two alternative scenarios when the AGN does not restart its radio activity, but instead (b) fully ceases the jet production or (c) is subject to increased activity (by e.g. undergoing a new accretion event, i.e. activity amplification). In the case of the jet switching off (b), and again invoking light travel arguments, we may observe FRII morphology on the far side, while the hotspot on the near side may have already faded away leaving behind only diffuse low level emission of a typical FRI morphology. Both scenarios, (a) and (b) assume the radio galaxy initially {has} an FRII morphology. For scenario (c) radio galaxy can be init{ia}lly of either FRI or FRII morphology, where for the former we will observe a reversed structure to the one of scenarios (a) and (b).

\subsubsection{Timescales}

In the absence of the re-acceleration of electrons, the hotspots will fade away on timescales of order \mbox{$\sim10^4-10^5$~yrs}, while the diffuse lobe emission will be slowly radiating away for $>10^7$~yrs. The light travel time arguments of the source morphological asymmetry require modulation of activity on timescales of \mbox{$\sim 10^5 - {\rm few} \times 10^6$~yrs} for physical scales of 100~kpc -- 1~Mpc. Hydrodynamical simulations of supermassive black hole temporal evolution \cite[e.g.][]{2011ApJ...737...26N,2014ApJ...789..150G} predict very chaotic accretion, with significant intermittency in accretion rate on a range of timescales, including bursts within a single accretion event (on orders of $10^6$~yrs) and long-term activity intermittency with multiple accretion events (activity with timescales of $\sim10^8$~yrs, separated by quiescent times of few~$\times 10^7$~yrs). {Timescales of $\sim10^5$~yrs for the typical AGN phase (optical and X-ray regimes), and so the timescales of the variability of the supermassive black hole accretion rates, has been also suggested by \cite{2015MNRAS.451.2517S}.} Such timescales are {in} agreement with the differential light travel arguments put forward in \S\ref{sec:nature}.


\begin{figure}
\includegraphics[angle=270,width=86mm]{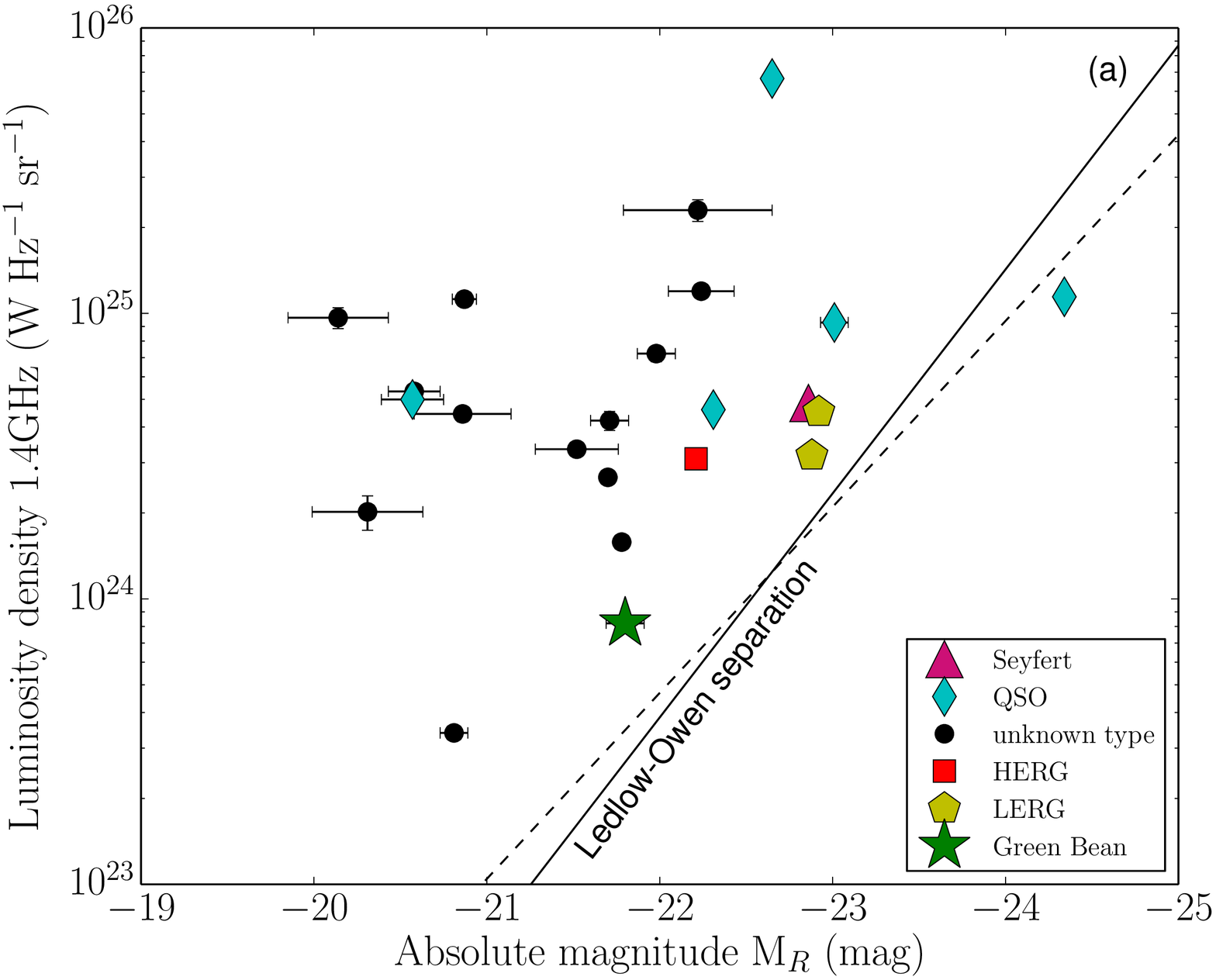}
\includegraphics[angle=270,width=86mm]{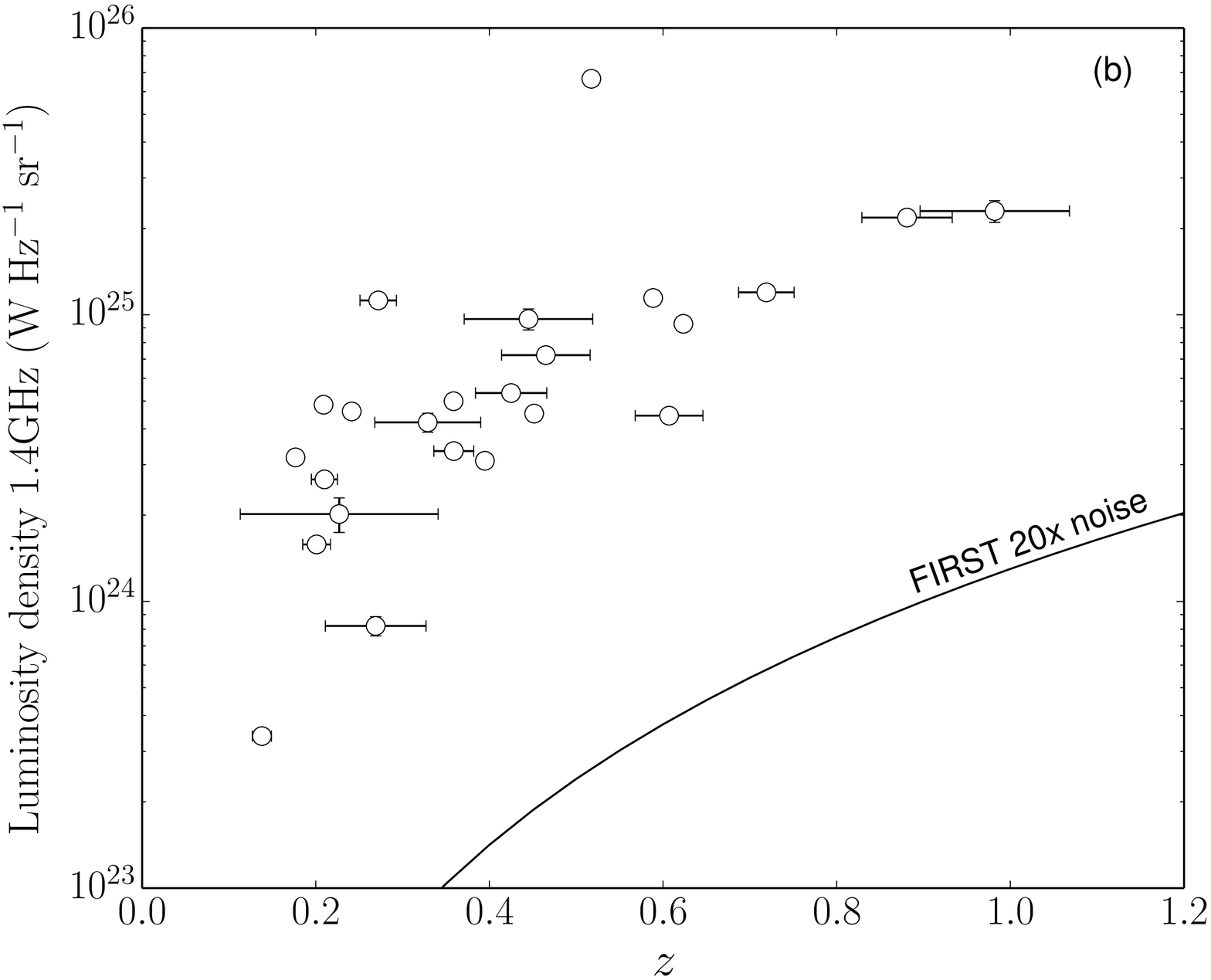}
\includegraphics[angle=270,width=86mm]{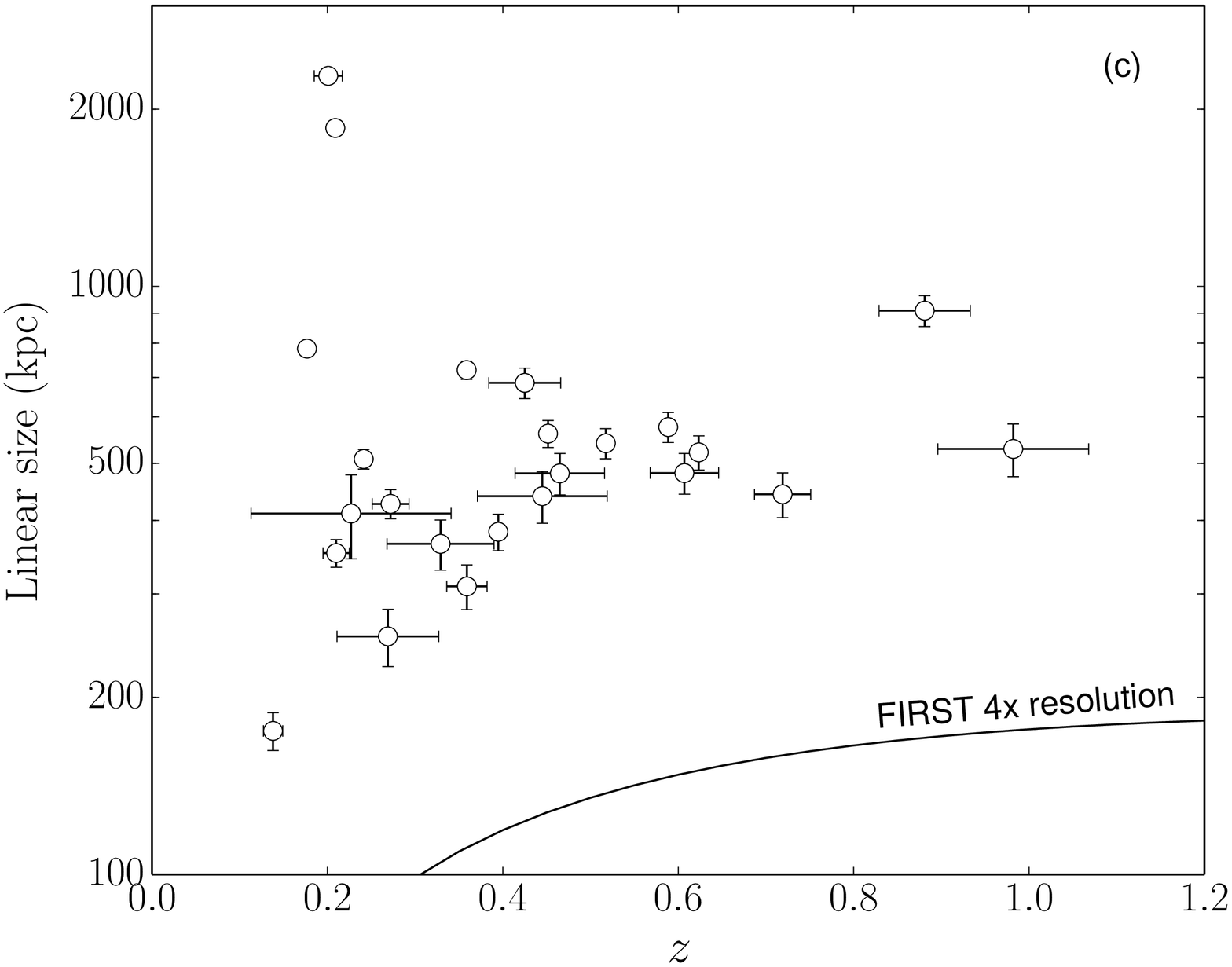}
\caption{(a) Ledlow-Owen diagram for the RGZ HyMoRS candidates. The solid line indicates the separation of FRI and FRII sources proposed by \protect\cite{1996AJ....112....9L}, and the dashed line the separation of FRIs and FRIIs updated by \protect\cite{2001A&A...373..100G}. (b) Luminosity density -- redshift distribution. The luminosity density equal to $20\sigma$ FIRST noise level (3~mJy) is drawn for reference. (c) Linear size -- redshift distribution. The linear size equal to $4\times$ FIRST resolution ($20''$) is drawn for reference. For discussion see \S{\ref{sec:disc-radio}}.}
\label{rys:radio-opt}
\end{figure}


\subsection{HyMoRS as quasars}
\label{sec:disc-quasars}

Interestingly, {6} of {9} candidates with spectroscopic observations (20\% of all our candidate HyMoRS) exhibit quasar properties. There is now a wealth of research that have shown that quasars are most likely triggered by mergers \cite[e.g.][]{1984AJ.....89..958H,2008ApJ...677..846B,2008ApJ...674...80U}, have close companions \cite[e.g.][]{1995Natur.376..150D} and may be residing in rich environments (\citeauthor{1991ApJ...371...49E} \citeyear{1991ApJ...371...49E}, \citeauthor{1993ApJ...411...43Y} \citeyear{1993ApJ...411...43Y}, but cf. \citeauthor{2000MNRAS.316..267W} \citeyear{2000MNRAS.316..267W}, \citeauthor{2001MNRAS.321..515M} \citeyear{2001MNRAS.321..515M}). We now know that radio loud quasars may be of both FR~II \cite[][and references therein]{1991ApJ...371...49E,1993ApJ...411...43Y} and FR~I radio morphology \cite[][]{2007MNRAS.381.1093H}. \cite{2007MNRAS.381.1093H} postulated that powerful radio sources may give rise to both radio morphological classes depending on the density of the environment in which the radio source expands. Confirmed HyMoRS whose hosts are quasars include J1348+286 \cite[][]{2006A&A...447...63G} and 1004+130 \cite[][]{2000A&A...363..507G}, while the host of the archetypal HyMoRS, NGC~612 \cite[][]{2000A&A...363..507G}, is a Seyfert. The existence of HyMoRS quasars suggests that their environments may be very asymmetric. Furthermore, at least one of our candidates resides in a cluster evironment; RGZ~J122425.8+02031{0} is its cluster BCG. This further advocates the impact of rich environments on the radio morphology of HyMoRS.

On the other hand, the existence of HyMoRS green bean galaxies, which is rare on both accounts [there are only $\lesssim20$ HyMoRS not including {candidates in this} paper \cite[e.g.][]{1996MNRAS.282..837S,2000A&A...363..507G,2006A&A...447...63G,2011BASI...39..547P,2017MNRAS.tmp..211D}, and 22 GBGs known to date \cite[][]{2013ApJ...763...60S,2015MNRAS.449.1731D,2016MNRAS.463.1554S}] suggests that modulation of central engine activity may be a non-negligible factor in the formation of hybrid radio morphology for at least some of the HyMoRS. The GBGs are ionisation echoes of quasars experiencing a high activity episode and a subsequent rapid shut down. These galaxies are associated with shock or ionisation fronts, including shocks of collimated jets heating the interstellar medium \cite[][]{2013ApJ...763...60S,2015MNRAS.449.1731D}. We refer the reader to Banfield et al. ({\it in prep}) on {the} detailed analysis of GBGs and their radio properties. 

It has been proposed that the AGN activity in GBGs has shut down within the last $10^4-10^5$~yrs \cite[][]{2013ApJ...763...60S,2016MNRAS.463.1554S}. RGZ~J123300.{2}+06032{5} has no clear radio core detected down to {the} $3\sigma$ upper limit of $0.5$~mJy, what suggests that the activity of the host is at most at low levels, if present at all. Furthermore, the projected linear extent of the RGZ~J123300.{2}+06032{5} radio structure of 255~kpc indicates that its radio jets were launched at least $4\times 10^5$~yrs ago, wh{ich} might have coincided, or been directly related to, the high activity episode of the AGN immediately preceding the shut down. We suggest that RGZ~J123300.{2}+06032{5} is a plausible example of the formation of HyMoRS through the amplification of the AGN activity, which we propose as an additional possible mechanism in this paper. This complements the central engine activity argument (activity cessation) put forward by \cite{2012AA...544L...2M,2012A&A...545A.132M}.

\subsection{Radio properties of HyMoRS and \\the Ledlow-Owen diagram}
\label{sec:disc-radio}

Overall, our candidate HyMoRS are moderately powerful radio galaxies. In Figure~\ref{rys:radio-opt}a we plot the locations of the candidates in the Ledlow-Owen diagram. The Ledlow-Owen diagram \cite[][]{1994ASPC...54..319O,1996AJ....112....9L} suggest {a} dependence of radio morphology on the absolute magnitude of the host of the radio galaxy. The dichotomy in radio morphology has been related to the optical brightness of the host through parameters such as black hole mass, pressure, accretion rate as all of these parameters can affect the jet propagation \cite[]{1995ApJS..101...29B,2001A&A...373..100G,2001A&A...379L...1G,2007A&A...470..531W, 2012AJ....144...85S}. 

In simple terms one can also interpret the Ledlow-Owen separation line between FRIs and FRIIs as a change in intrinsic power of a radio galaxy, where FRIs are typically less powerful than FRIIs and thus their observed radio luminosity density is lower \cite[but see ][for details on degeneracies between radio luminosity density and kinetic luminosity of jets, which lead to {a} non-straightforward mapping between these two parameters]{2012MNRAS.424.2028K}. One could naively expect, therefore, that the weaker jets are more easily disrupted when expanding in a non-uniform external medium. Numerical simulations of \cite{2008A&A...491..321M}, who attempted to model the distruption of only one of the twin jets by assuming a density jump in the external medium, show that weak jets ($10^{36}$~W) can indeed efficiently form HyMoRS. However, they also show that for a right set of parameters (e.g. jet speed and density, and density ratio between the external dense medium and jet) hybrid radio morphology can also be formed in radio galaxies with powerful jets ($10^{39}$~W). {The fact that our candidate HyMoRS are located in the same region as powerful FRIIs (Figure~\ref{rys:radio-opt}a) suggests that the latter may be  happening for at least some of the HyMoRS.} The range of the observed radio luminosity densities of the candidate HyMoRS (Figure~\ref{rys:radio-opt}b), with a median $L_{\rm median}={4.7}\times10^{24}$~W~Hz$^{-1}$~sr$^{-1}$ and a range of ${3.4 }\times10^{23}-6.7\times10^{25}$~W~Hz$^{-1}$~sr$^{-1}$, seem to be in agreement with the results of \cite{2008A&A...491..321M}.

Since the seminal study of \cite{1996AJ....112....9L}, the classification of radio galaxies based on their optical spectra has been attracting more attention. Specifically, radio galaxies can be classified as either high excitation (quasar-mode) or low excitation \cite[jet-mode; e.g. ][]{1994ASPC...54..201L,2012MNRAS.421.1569B}. The FRI/FRII classification cannot be directly mapped onto the HERG/LERG classification: while HERGs are typically powerful FRIIs, LERGs can be {either} FRIs {or} low power FRIIs. The low power FRII type radio galaxies occupy the FRI region in the Ledlow-Owen diagram \cite[see e.g.][]{2009AN....330..184B,2017MNRAS.tmp...25M}. In fact, based on Figure~5 in \cite{2017MNRAS.tmp...25M} it may seem that the separation occurs between the excitation classes rather than the morphological types of radio galaxies. {A s}imilar conclusion has been recently reached by \cite{2017arXiv170303427C} who used the same base data sample as \cite{2017MNRAS.tmp...25M}, but see discussion in \cite{2014Ap&SS.353..233S} {on the validity} and \cite{2015Turner} {on} tightness of {and environmental impact on} the Ledlow-Owen correlation. Here, we find that {our} HyMoRS {candidates} can be both low- and high-excitation radio galaxies, although a study with larger number statistics is needed to understand how common each is.

Our sources display a range of linear sizes, from 175~kpc to Mpc-scales, with two sources classified as giant radio galaxies {(Figure~\ref{rys:radio-opt}c)}. The apparent scarcity of low luminosity and small linear size sources in our selection {may be} artificial, however; simply, we are less likely to select faint ($\lesssim20$~mJy) or small angular size sources ($\lesssim30-40$~arcsec) because the morphological classification becomes more ambiguous in those instances. {This bias may also affect the distribution of HyMoRS in the Ledlow-Owen diagram (Figure~\ref{rys:radio-opt}a)}. Despite this, our current results already indicate a large diversity in both the type of host in which HyMoRS {may} reside and their radio properties. This suggests that, in principle, any active galaxy may be host to a hybrid morphology radio galaxy, and the morphology can be created by both the environment and central engine activity modulation.


\section{Summary \& Conclusions}
\label{sec:summary}

We present the first 25 new candidates of hybrid morphology radio sources (HyMoRS) drawn from the international citizen science project, Radio Galaxy Zoo, and its online discussion forum, the RadioTalk. For all the candidates, we provide mid-IR and optical hosts, redshifts, and radio and optical properties (luminosities, sizes). This is the first time such a large sample of candidate HyMoRS, with {ancillary} data on their hosts, has been collated. We discuss possible scenarios of the formation of hybrid morphology of radio galaxies, including:
\begin{itemize}
\item[(i)] non-uniform environments,
\item[(ii)] cessation or amplification of the activity of the central black hole, and
\item[(iii)] Doppler boosting.
\end{itemize}
Detailed radio spectral and polarimetric analyses are needed to distinguish between these scenarios for each HyMoRS, but based on the available data, we postulate that HyMoRS are a diverse class of objects that in principle can be formed by any of these mechanisms and can be hosted by any active galaxy.

We cross-match our serendipitously selected candidate HyMoRS with the upcoming Radio Galaxy Zoo DR1 catalog to quantify the accuracy with which the citizen scientists identify and classify these complex sources. We find the citizen scientists identify the correct mid-IR host in at least 80\% of cases, and correctly identify all radio components of each radio galaxy in 80\% of cases. For radio galaxies of angular sizes smaller than $115''$ the success rates are 90\% for the mid-IR host identification, and 100\% for the radio component association. These results are very promising for future blind selection of candidate HyMoRS from the RGZ catalogs.  

Given the rarity of these sources, and sheer volume of the data, we intend to pre-select all candidate HyMoRS from the FIRST survey using the Radio Galaxy Zoo project, for high-resolution continuum and polarimetric follow-up observations, and for efficient construction of future large HyMoRS samples. Deep follow-up studies of seven presented here candidates are currently in progress (Kapinska et al. {\it in prep}). Future high-resolution all-sky surveys, such as NRAO VLA Sky Survey\footnote{\tt https://science.nrao.edu/science/surveys/vlass} (VLASS)  which will have twice as high resolution as FIRST, will be of a particular value and great efficiency in the confirmation of the candidates. We highlight, however, the need for multi-resolution and/or multi-frequency radio data for at least some of the sources.


\section*{Acknowledgments}

A.D.K. thanks P.A. Curran for constant encouragement in achieving academic goals. 

This publication has been made possible by the participation of more than  11,000 volunteers in the Radio Galaxy Zoo Project. Their contributions are acknowledged at http://rgzauthors.galaxyzoo.org. { We thank the following volunteers in particular for their comments on the manuscript or active search for candidate RGZ HyMoRS on RadioTalk: Jean Tate, Tsimafei Matorny, Victor Linares Pag{\'a}n, Christine Sunjoto, Leonie van Vliet, Claude Cornen, Sam Deen, K.T. Wraight, Chris Molloy, and Philip Dwyer.} Along with the contribution of the Radio Galaxy Zoo volunteers, we also acknowledge A. Kapadia, A. Smith, M. Gendre, S. George, E. Paget, R. Simpson, and C. Snyder who have made contributions to the project. The development of Radio Galaxy Zoo was supported by a grant from the Alfred P. Sloan foundation. 
A.D.K. and J.K.B. acknowledge financial support from the Australian Research Council Centre of Excellence for All-sky Astrophysics (CAASTRO), through project number CE110001020. 
S.S.S. thanks the Australian Research Council for an Early Career Fellowship (DE130101399). 
Partial support for this work for K.W.W. and L.R. was provided the U.S. National Science Foundation grant AST-1211595 and AST-1714205 to the University of Minnesota. 
K.S. acknowledges support from Swiss National Science Foundation Grants PP00P2\_138979 and PP00P2\_166159. 
F.d.G. is supported by the VENI research programme with project number 1808, which is financed by the Netherlands Organisation for Scientific Research (NWO). 
H.A. benefitted from grant DAIP 980/2016-2017 of Univ. of Guanajuato. 

This work made use of the FIRST and NVSS National Radio Astronomy Observatory (NRAO) Very Large Array (VLA) surveys. NRAO is a facility of the National Science Foundation operated under cooperative agreement by Associated Universities, Inc. This publication makes use of data products from the Wide-field Infrared Survey Explorer, which is a joint project of the University of California, Los Angeles, and the Jet Propulsion Laboratory/California Institute of Technology, funded by the National Aeronautics and Space Administration. Funding for the SDSS and SDSS-II has been provided by the Alfred P. Sloan Foundation, the Participating Institutions, the National Science Foundation, the US Department of Energy, the National Aeronautics and Space Administration, the Japanese Monbukagakusho, and the Max Planck Society, and the Higher Education Funding Council for England. The SDSS Web site is http://www.sdss.org/. This research has made use of the NASA/IPAC Extragalactic Database (NED) which is operated by the Jet Propulsion Laboratory, California Institute of Technology, under contract with the National Aeronautics and Space Administration. This research used TOPCAT -- Tool for OPerations on Catalogues And Tables { \cite[][]{2005ASPC..347...29T}. Some of the data used in this paper were obtained from the Mikulski Archive for Space Telescopes (MAST). STScI is operated by the Association of Universities for Research in Astronomy, Inc., under NASA contract NAS5-26555. Support for MAST for non-HST data is provided by the NASA Office of Space Science via grant NNX09AF08G and by other grants and contracts.}

\bibliographystyle{mn2e} 
\bibliography{./mybib}

\label{lastpage}

\end{document}